\documentclass[]{iopart}

\usepackage[colorlinks]{hyperref} 
\usepackage{graphicx}
\include{Qcircuit}

\usepackage{iopams}

\date{June 30^{\mbox{\footnotesize th}}, 2006}

\begin{document}
\title{Qudit surface codes and gauge theory with finite cyclic groups}

\author{Stephen S. Bullock}
\address{IDA Center for Computing Sciences,
17100 Science Drive, Bowie, MD 20715-4300 USA}\ead{ssbullo@super.org}
\author{Gavin K. Brennen}
\address{Institute for Quantum Optics and Quantum Information,
Techniker Str. 21a, 6020, Innsbruck, Austria}\ead{gavin.brennen@uibk.ac.at}

\begin{abstract}
Surface codes describe quantum memory stored as a global property 
of interacting spins on a surface.  The state space is fixed by 
a complete set of quasi-local stabilizer operators and the 
code dimension depends on the first homology group of the surface 
complex.  These
code states can be actively stabilized by measurements or, alternatively, 
can be prepared by cooling to the ground subspace of a 
quasi-local spin Hamiltonian.
In the case of spin-1/2 (qubit) lattices, such ground states have 
been proposed as 
topologically protected memory for qubits.  We extend these 
constructions to lattices
or more generally cell complexes with qudits, either of prime
level or of level $d^\ell$ for $d$ prime and $\ell \geq 0$, and therefore under
tensor decomposition, to arbitrary finite levels.  
The Hamiltonian describes an
 exact $\mathbb{Z}_d\cong\mathbb{Z}/d\mathbb{Z}$ gauge theory 
whose excitations correspond
to abelian anyons.  We provide protocols for qudit storage and retrieval 
 and propose an interferometric verification of topological order by measuring 
 quasi-particle statistics.
\end{abstract}

\maketitle


\section{Introduction}
There is a rich history to the study of topologically ordered states 
of matter.  Such states are defined by the property that all physical
correlation functions are topological invariants.  In the field of condensed 
matter, these states have been proposed as ground states of models
for high temperature superconductors and for fractional quantum Hall 
states \cite{Wen}.
Furthermore, it has been demonstrated that such order can arise as 
a low energy property of hard core bosonic spin lattice models.  In 
contrast to 
the familiar situation with spontaneous symmetry breaking, here the ground 
states exhibit more symmetry than the microscopic equations of motion. 
 It has been suggested that such emergent properties may model
gauge fields and particles found in nature \cite{Levin:05}.  
In the field of quantum information it was shown by Kitaev 
\cite{Kitaev} that ground states of Hamiltonians which can be expressed
as a sum over quasi-local stabilizer operators provide for 
topologically protected qubit memories.
These states are referred to as surface codes.  They are robust to arbitrary 
quasi-local perturbations and have abelian
anyonic excitations.  In order to perform universal
fault-tolerant quantum processing, it is necessary to use non abelian 
anyonic excitations that transform under an appropriate group \cite{Kitaev}.  
From the algorithmic point of view attempts have also been made to understand 
quantum computing in terms of nonabelian anyon operations \cite{jones}.
Finding suitable microscopic lattice models that provide for 
universal quantum computation
is an area of active research \cite{quantumHall}.

This survey attempts to exhaust the topic of surface codes 
for topologically protected qudit memories.  While not as 
powerful as fault tolerant models with non abelian anyons,
these models offer a new perspective on non-local encoding 
of quantum information and give us insight into microscopic realizations of
lattice gauge theories.  
Surface codes for two level systems \cite{Kitaev} are by now
well understood.  Their implications for error-resistant quantum computer
memories have also been considered \cite{dennis}.  In the error-correction
context, the topologically
ordered eigenstates may be understood as a particular case of
quantum stabilizer codes (e.g. \cite{Gottesman}.)  The error lengths
of the resulting stabilizer codes are not exceptional, and only rarely do
anyonic systems appear in classifications of near-optimal quantum codes.
(Optimality in this sense refers to minimizing the number of code-qubits
against the number of errors a code may correct.)
Yet all the error correction operations are local upon the lattice
in which the quantum data is stored, which might improve scalability.
Moreover, an aside to an argument focused on deriving a famous stabilizer
code from the topology of the real projective space in fact demonstrates
that a qubit lattice is not required \cite{FreedmanMeyer}.  Rather,
a two-complex (see e.g. \cite{topology}) 
suffices, where a two-complex is a generalization of
a graph in which discs are also allowed with edge boundaries.
On the physical system which places a qubit on each edge of a 
(cellular or simplicial) two-complex 
$\Gamma$, there exists a Hamiltonian whose topologically
ordered (stabilizer-code) groundstates are parametrized by the first
homology group of the complex with bit-coefficients:
$H_1(\Gamma,\mathbb{F}_2)$.  The Hamiltonian is a sum of vertex and
edge terms which are proportional to either
tensors of Pauli $Z$ operators around qubits on edges
adjacent to a vertex or are proportional to tensors of
$X$ operators on edges bounding a face of the complex.

For some time the existence of stabilizer
codes over qudits ($d$ prime) have been known \cite{Gottesman}.
Yet only recently have results on the topic become as strong as
those applicable in the bit case, including estimates
of optimal code-lengths etc. (\cite{TA&M},
see also \cite{hostens}.)  Moreover, extensions
to prime-power ($d^\ell$) level qudits (actually qu$d^\ell$its)
have also been found, so that tensors provide a stabilizer
formalism for all finite-level systems.  In this work,
we exploit the new stabilizer formalism to construct codes
on a two-complex whose edges carry prime-$d$-level
qudits, and we also outline the extension to $d^\ell$-level qudits.
The associated ground states are parametrized by
$H_1(\Gamma,\mathbb{F}_{d^\ell})$, where the coefficient field
is viewed as an abelian group under addition.  This requires few
new ideas, although care must be taken with sign conventions
which were vacuous in the earlier work on $\mathbb{F}_2$-coefficients.
Thus, after tensoring we have constructed surface codes
with qudits for arbitrary finite $d$ placed on the edges
of a generic orientable two-complex $\Gamma$.  Recent work by 
Bombin and Martin-Delgado \cite{Bombin} investigates classical 
and quantum homological error correction codes.  They construct a
class of surface codes for qudits which asymptotically saturates the
maximum coding rate and provide several example encodings
on various two complexes.  Here we do not address the issue of coding
efficiency.  Rather we concentrate on explicit
constructions of Hamiltonians that support qudit surface codes
in their ground eigenstates and describe how one might 
encode and decode therein.

The manuscript is intended to be self-contained.  Thus, 
\S\ref{sec:stabilizer} opens by reviewing
some the required facts on stabilizer codes.  In order to aid
readers less interested in the general case, \S\ref{sec:prime}
treats prime-$d$ level encoding on surfaces separately.
Methods for encoding, decoding, and stabilizer measurements
are given in \S\ref{sec:memory}.  Extensions to the case of prime
power qudit encodings are given in \S\ref{sec:extensions}.
Errors in our model correspond to 
low lying excitations in the Hamiltonian whose superselection sectors 
may be viewed as massive particles on the underlying cellulation.  
In \S\ref{anyons}
it is shown that our model reproduces a 
$\mathbb{Z}/d\mathbb{Z}$ gauge theory
where errors are described by particle anti-particle pairs of 
change/flux dyons.  We propose an interferometer circuit for measuring the
the statistics of these quasiparticles.  We conclude with a summary and
some. 

\section{Qudit Stabilizer Codes}
\label{sec:stabilizer}

We next review stabilizer codes \cite{Gottesman,TA&M}.  This section
focuses on the case of qudits with a prime number of levels.  
The first subsection recalls the definition
and a basic technique.
The next subsection generalizes a well known construction
from bits to dits.

\subsection{Stabilizers and groundstates}

Let $d$ be a prime number, and consider the qudit
state space $\mathcal{H}(1,d)=\mathbb{C} \ket{0} \oplus \cdots \oplus
\mathbb{C} \ket{d-1}$, with a pure state of $n$ qubits being a ket
within $\mathcal{H}(n,d)=\mathcal{H}(1,d)^{\otimes n}$.  A possible
generalization of the Pauli operators on $\mathcal{H}(1,d)$ would be
to consider the group generated by the following unitary matrices:
\begin{equation}
\begin{array}{lcll}
X \ket{j} &  = & \ket{j+1 \mbox{ mod }d} & \\
Z \ket{j} & = & \xi^j \ket{j}, & \mbox{ for }\xi = \mbox{exp}(2 \pi i/d) \\
\end{array}
\end{equation}
These are not Hermitian unless $d=2$.  The qudit Pauli-tensor
group, say $\mathcal{P}(n,d) \subsetneq U[\mathcal{H}(n,d)]$, is the group
of unitary matrices generated by $n$-fold tensors of elements of
$\{I_d,X,Z\}$.

We might be more explicit in the description of $\mathcal{P}(n,d)$.
First, for $n=1$, label the multiplication in
$\mathbb{F}_d$ to be a dot-product.  Then
$ Z^b X^a= \xi^{a \bullet b}X^a Z^b$.  More generally, for dit-strings
$a,b \in (\mathbb{F}_d)^n$, we use $X^{\otimes a}$ and $Z^{\otimes b}$
to abbreviate $X^{a_1} \otimes X^{a_2} \otimes \cdots \otimes X^{a_n}$
and similarly $Z^{\otimes b}$ for
$Z^{b_1} \otimes Z^{b_2} \otimes \cdots \otimes Z^{b_n}$.
For the $n$-entry dot-product with values in $\mathbb{F}_d$, we
have 
$Z^{\otimes b} X^{\otimes a}= \xi^{a \bullet b} X^{\otimes a} Z^{\otimes b} $.
Thus explicitly 
\begin{equation}
\mathcal{P}(n,d)\ = \ \{ \xi^c X^{\otimes a} Z^{\otimes b} \; ; \;
a,b \in (\mathbb{F}_d)^n, c \in \mathbb{F}_d \big\}
\end{equation}
The qudit stabilizer groups are subgroups $G \subseteq
\mathcal{P}(n,d)$.  The code subspace of such a stabilizer group
is the joint $+1$ eigenspace of all $g \in G$.

Of course, such joint eigenspaces might well be trivial.
Yet a standard argument shows that they are nontrivial
in certain cases.  This technique is so fundamental to
stabilizer code manipulation that we wish to highlight it;
it will be used several more times in the course of the work.
While actually an elementary technique from representation theory,
it has also featured prominently in the quantum computing
literature \cite{Hallgren}.

\noindent
{\bf Stabilizer code projectors:}
\emph{The sum of unitary maps
$\pi=(\# G)^{-1}\sum_{g \in G} g$ is a projection
onto the code-subspace.}  
We present the argument.  First, $\pi^2$ is the identity
map since $\pi g = \pi$ for any $g \in G$.  Second, $\pi=\pi^\dagger$
since adjoints are inverses in the unitary group.  Hence,
either $\pi$ is a projection or $-1$ is an eigenvalue of $\pi$.
Yet $I_{d^n}$ is a summand, so the complex inner product precludes
$\pi \ket{\psi}=-\ket{\psi}$ for a nonzero $\ket{\psi}$.
Now split
$\mathcal{H}(n,d)=V_1 \oplus V_2 \oplus \cdots \oplus V_\ell$ into 
irreducible orthogonal unitary subrepresentations of
$G$.  For each $V_j$, the image under $\pi$ and its orthogonal
complement form a decomposition of $V_j$.  Thus
by irreducibility, $\pi$ either preserves a $V_j$ or $\pi V_j=0$.
Clearly the former holds for any irrep 
(i.e. irreducible representation)
within the code subspace of $G$.
On the other hand, if $\langle \psi | g | \psi \rangle \neq 1$ for
some $g$, then the latter holds.

As a remark, irreps within the code subspace of $G$ must be
one-dimensional and are also known as \emph{trivial representations}.
As a second remark,  the code subspace is nonzero
iff $\mbox{Trace}(\pi) \neq 0$ iff $(G \cap \{ \xi^j I_d\})=\{I_{d^n}\}$.

In the Hermitian case ($d=2$,) 
it is standard that all eigenvalues of group elements
are $\pm 1$, so that a suitable Hamiltonian for which the
code space is the groundstate is $-\pi$.  For general $d$, the
eigenvalues lie within the unit circle, so that
$-1$ is still the least possible real part.  
Also, $g^\dagger \ket{\lambda}=(1/\lambda) \ket{\lambda}
= \overline{\lambda} \ket{\lambda}$ since $g^\dagger=g^{-1}$.
Thus, one may place the
qudit code subspace into the groundstate of a Hamiltonian by adjusting
each summand of $\pi$ with a Hermitian conjugate:
$H=\sum_{g \in G} -(g + g^\dagger)$, so that the eigenvalues of the summands
are then $- 2 \mbox{Re}[\mbox{spec}(g)]$.

\subsection{Quantum circuits for qudit stabilizer measurements}
\label{sec:stab_measure}

Given an $n$-qudit system, it is important for purposes of
error-correction to be able to test whether or not a state
$\ket{\psi}$ lies within the stabilizer code of some
$G=\langle \{g_j\} \rangle \subseteq \mathcal{P}(n,d)$.
It suffices to test whether $\ket{\psi}$ is a $+1$ eigenvector
of each generator $g_j$.  We sketch quantum circuits 
which achieve such a measurement.

Let $\mathcal{F}_d=d^{-1/2}
\sum_{j,k=0}^{d-1} \xi^{jk} \ket{j}\bra{k}$ be the qudit Fourier
transform.  Considering eigenkets, $\mathcal{F}_d^\dagger X \mathcal{F}_d
=Z$.  Now the number operator ${\bf n}=\sum_{j=0}^{d-1}
j \ket{j}\bra{j}$ suffices to infer
the eigenvalue of $Z$ and project into the appropriate eigenstate.
As a circuit, we might denote a number operator measurement
with the $Z$ symbol, one of several common conventions in the
qubit case:
\[
\Qcircuit @C=1em @R=2em
{
& \meter & \cw \\
}
\]
Determination of the $X$ eigenstate may be accomplished by
\[
\Qcircuit @C=1em @R=2em
{
& \qw & \gate{\mathcal{F}_d^\dagger} & \meter & \cw
}
\]
Similarly, there is some one-qudit unitary which will
diagonalize any $X^a Z^b \in \mathcal{P}(1,d)$, usually
not a Fourier transform.  Yet using the diagonalization
and a number operator one may infer an eigenstate.

For $Z^{\otimes k}$ and $X^{\otimes k}$, we suggest using addition
gates along with a qudit ancilla.  We will denote
$\ket{j,k} \mapsto \ket{j,(j+k)\mbox{ mod }d}$ by a typical control
bullet with the target (in the formula second) line holding a $+$ gate.
The the following construction of $Z^{\otimes 2}$ generalizes for
$Z^{\otimes k}$:
\[
\Qcircuit @C=1em @R=2em
{
& \lstick{\ket{0}} & \gate{+} & \gate{+} & \meter & \cw \\
& \qw & \ctrl{-1} & \qw & \qw & \qw \\
& \qw & \qw & \ctrl{-2} & \qw & \qw 
}
\]
For $Z^{\otimes k} \ket{j_1,j_2,\ldots,j_k}=
\xi^{j_1+\cdots +j_k} \ket{j_1,j_2,\ldots,j_k}$, and we have
placed 
$\ket{(j_1+\cdots + j_k) \mbox{ mod }d}$ on the ancilla line
before the number operator is applied.  Note that $Z\otimes Z^{-1}$
results by replacing one of the modular addition gates above
with modular subtraction.  Powers of operators 
are measured by multiple appications of the sum gate appropriately.
Finally, 
$(\mathcal{F}_d^\dagger)^{\otimes k} X^{\otimes k} \mathcal{F}_d^{\otimes k}
=Z^{\otimes k}$, so that the following diagram for $X^{\otimes 2}$
extends:
\[
\Qcircuit @C=1em @R=2em
{
& \lstick{\ket{0}} & \gate{+} & \gate{+} & \meter & \cw \\
& \gate{\mathcal{F}_d^\dagger} & \ctrl{-1} & \qw & \gate{\mathcal{F}_d} 
& \qw \\
& \gate{\mathcal{F}_d^\dagger} & \qw & \ctrl{-2} & \gate{\mathcal{F}_d} & \qw 
}
\]
Using similarity transforms by qudit Fourier transforms, we may similarly
achieve $X\otimes Z \otimes Z \otimes X$ etc.  Yet more generally, the
comment on existence of diagonalizations above produces circuits for
arbitrary elements $g \in \mathcal{P}(n,d)$. 
\section{Homologically Ordered Groundstates for Prime Qudits}
\label{sec:prime}

It is typical to place topological orders on explicit planar
or spacial lattices of spin-$j$ particles,
e.g. square, triangular, hexagonal, Kagome, etc.
An alternative was presented in Freedman and Meyer's
derivation of certain error-correcting codes
of Shor and LaFlamme \cite{FreedmanMeyer}.  
Namely, qubits could be placed on the edges of a two-complex $\Gamma$,
and an appropriate Hamiltonian would have the dimension of its
degenerate groundstate eigenspace equal to the 
number of classes within $H_1(\Gamma,\mathbb{F}_2)$.  We next extend
this construction to prime-level qudits; the task is mainly to
keep track of sign conventions which are vacuous in $\mathbb{F}_2$.
We then check that the groundstate eigenspace
is similarly spanned by kets associated to elements of
$H_1(\Gamma,\mathbb{F}_d)$, by applying stabilizer-code
techniques.

\subsection{Cellular Hamiltonians}

Label $\mathcal{V}$ to be the vertices of $\Gamma$,
$\mathcal{E}$ to be the edges, and $\mathcal{F}$ to be the
faces.  We also require properties that hold if
$\Gamma$ is a cellulation of an orientable, compact, connected surface.
Specifically, each edge has a boundary of exactly two vertices
and each face has an orientation according to
which each edge lies in the boundary of two faces
with the edge taking opposite orientations in the boundary
of each face.  Finally, $\Gamma$ is finite and
$H_2(\Gamma,\mathbb{F}_d)$ is a copy of $\mathbb{F}_d$
spanned by $[\Gamma]$,
the sum of all faces with their orientation according to
$\Gamma$.  

We briefly review the appropriate homology.
Label the chain sets to be formal sums of
vertices, edges, and faces respectively:
\; \; $C_0(\Gamma,\mathbb{F}_d)=
\mbox{span}_{\mathbb{F}_d}(\mathcal{V})$,
$C_1(\Gamma,\mathbb{F}_d)=
\mbox{span}_{\mathbb{F}_d}(\mathcal{E})$, and
$C_2(\Gamma,\mathbb{F}_d)=\mbox{span}_{\mathbb{F}_d}(\mathcal{F})$.  
We generally drop the $\Gamma$ and coefficient system,
which should be clear from context.
Since $\Gamma$ is a cell complex, there exist boundary operators
\begin{equation}
C_0 {\buildrel \partial \over \longleftarrow} \ 
C_1  \ {\buildrel \partial \over \longleftarrow} \ 
C_2  
\end{equation}
with $\partial^2=0$ \cite{topology}.  
For example, if an edge $e$ connects $v_1$
and $v_2$, say $e=[v_1,v_2]$, then
$\partial e = v_1 - v_2 = v_1 + (d-1) v_2$.  
Note that the definition of $\Gamma$ demands that edges
$e \in \mathcal{E}$ are images of $[0,1]$ within $\Gamma$,
and hence all edges are implicitly \emph{oriented}.
The coefficients further allow for $\mathbb{F}_d$-valued
multiplicities on each edge.
Since $\partial^2=0$,
we have $\mbox{ker}(\partial_1) \supseteq 
\mbox{image}(\partial_{2})$ for $\partial_j:C_j \rightarrow C_{j-1}$.
Thus we may define the $\mathbb{F}_d$ vector space
$H_1(\Gamma,\mathbb{F}_d)=\mbox{ker}(\partial_1)/\mbox{image}(\partial_2)$.
This first homology group is well known to be a topological invariant,
i.e. any topological space homotopic to that underlying
$\Gamma$ will produce an $H_1(\Gamma,\mathbb{F}_d)$ of 
the same dimension.  Homology elements
are represented by cycles, i.e. elements of the kernel of the boundary 
operator.  However, several elements might represent the same class,
differing by a boundary, i.e. an element of $\partial (C_2)$.

Recall that any Hamiltonian on $n$-qudits may be written
as a sum of tensor products of Hamiltonians (Hermitian matrices)
on each factor.  The degree of a summand in the tensor basis
is the greatest number of non-identity factors in any term.
A $k$-local Hamiltonian is a Hamiltonian whose degree is bounded
by $k$ in some decomposition.  The topologically ordered
Hamiltonians defined below are $k$-local for $k$ the maximum
of the valence of any vertex and the number of edges on any face.

Let $n=\# \mathcal{E}$, and consider placing a qudit on each
$e \in \mathcal{E}$.  
Again, each edge is the image of $[0,1]$ and is oriented (by $\Gamma$)
from one vertex to the other.  For the qudits associated with
each edge, the $\ket{1}$ excitation of the
edge will be implicitly associated to this orientation, while the
$\ket{d-1}$ state corresponds to the other.

On the associated physical system $\mathcal{H}(n,d)$,
let $X_e$ and $Z_e$ denote the operator applied to the qudit of
that edge with identity operators buffered into the remainder of
the tensor.  For each $v \in \mathcal{V}$, we define a Pauli-tensor
and vertex Hamiltonian by
\begin{equation}
\begin{array}{lcl}
g_v & = & \prod_{e=[\ast,v]} Z_e \prod_{e=[v,\ast]} Z_e^{-1} \\
H_v & = & -(g_v + g_v^\dagger) \\
\end{array}
\end{equation}
For some $U>0$, we then define the potential energy term of
a topologically ordered Hamiltonian by $H_\partial = U
\sum_{v \in \mathcal{V}} H_v$.

The notation $H_{\bf \partial}$
has been chosen for the following reason.  Suppose
that $\omega = \sum_{e \in \mathcal{E}} n_e e$ is a chain,
with each $n_e \in \mathbb{F}_d$.  There is an associated qudit
computational basis state, say $\ket{\omega}$, which is local
and places the qudit of each $e$ in state $\ket{n_e}$.
We claim that $\ket{\omega}$ is a groundstate of $H_\partial$
iff $\partial \omega=0$, i.e. $\omega$ is a cycle.
To see this, one verifies that $g_v \ket{\omega} = \xi^c \ket{\omega}$
where $\partial \omega = c v + \sum_{w \neq v} c_w w$.
Hence $\ket{\omega}$ is in the stabilizer $\langle \{ g_v \} \rangle
\subseteq \mathcal{P}(n,d)$ iff $\ket{\omega}$ is an eigenstate
of each $H_v$ of minimial (real) eigenvalue iff
$\ket{\omega}$ is in the degenerate groundstate 
eigenspace of $H_\partial$.

Strictly speaking, one should not refer to the groundstate of
$H_\partial$ as being topologically ordered.  Admittedly, groundstates
are of the form $\ket{\psi_g} = \sum \alpha_\omega \ket{\omega}$ for
$\omega$ a cycle, colloquially a loop of excited edges.  For $d>2$,
the edges must be properly oriented, and hitting every edge of a 
$Y$ junction is allowed if multiplicities are accounted for.
Yet the cycle subspace is not a topological invariant.  
Indeed, should $\Gamma$ be a cell complex, subdividing
$\Gamma$ by breaking each $2$-simplex (triangle) into several
subtriangles will generally increase the size of $\mbox{ker}(\partial_1)$,
although such a subdivision does not change the topology of the
underlying manifold.  Thus,
we next add a kinetic energy term to the potential, splitting the
degeneracy of $H_\partial$ and reducing to a final groundstate
capturing homology.

\begin{figure}
\begin{center}
\includegraphics[width=\columnwidth]{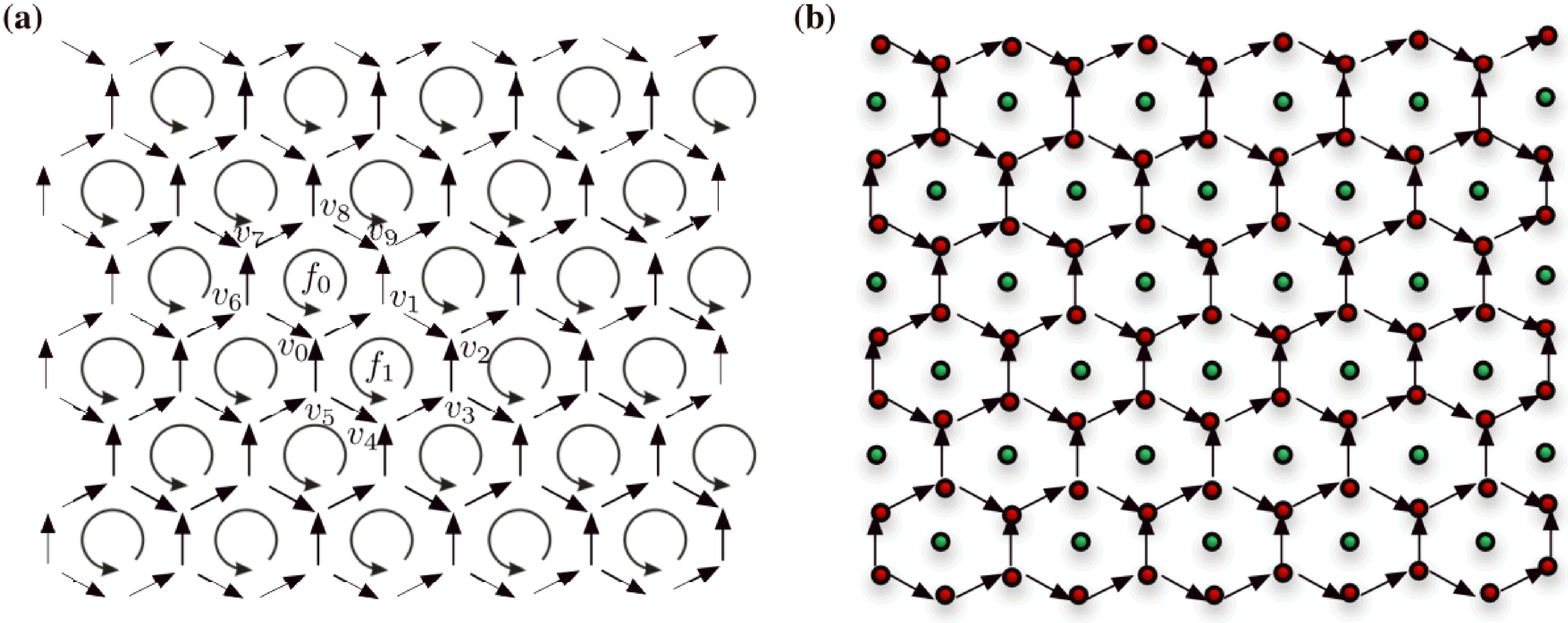}
\caption{\label{fig:1}  Cellulation of an orientable surface.  
Each system particle (qudit) is represented by an edge.  
Particle interactions occur between all edges that meet at a common 
vertex and all edges comprising a plaquette boundary.  (a)  In this example, 
physical qudits reside on the vertices of a Kagome' lattice on a torus such 
that the resultant cellulation is a honeycomb lattice on a torus.  Edge and 
face orientations are indicated.  For the vertices $v_0,v_1$ and 
faces $f_0,f_1$ the mutually commuting operators in the Hamiltonian are 
$g_{v_0}=Z_{[v_6,v_0]}Z_{[v_5,v_0]}Z^{-1}_{[v_0,v_1]}$, 
$g_{v_1}=Z_{[v_0,v_1]}Z^{-1}_{[v_1,v_9]}Z^{-1}_{[v_1,v_2]}$, 
$g_{f_0}=X_{[v_0,v_1]}X_{[v_1,v_9]}X^{-1}_{[v_8,v_9]}X^{-1}_{[v_7,v_8]}
X^{-1}_{[v_6,v_7]}X_{[v_6,v_0]}$,\ 
$g_{f_1}=X^{-1}_{[v_0,v_1]}X^{-1}_{[v_1,v_2]}X_{[v_3,v_2]}
X_{[v_4,v_3]}X_{[v_5,v_4]}X^{-1}_{[v_5,v_0]}$.  
(b)  Same cellulation with vertex (red) ancilla and face (green) ancilla.  
These can be used to perform local stabilizer checks
or to mediate many body
interactions between edges from physical $2$-local interactions as
described in \S\ref{secancstab}.}
\end{center}
\end{figure}

For each face $f$, the face Hamiltonian $H_f$ is defined
as follows.  Orient $f$ according to the orientation of the
manifold underlying $\Gamma$.  Label edges by
$\partial f = \sum_{k=1}^{p} o_k e_k$ for
$o_k \in \{1,d-1\}$.  Then we define
\begin{equation}
\begin{array}{lcl}
g_f & = & X_{e_1}^{o_1} X_{e_2}^{o_2} X_{e_3}^{o_3} \ldots
X_{e_p}^{o_{p}} \\
H_f & = & -(g_f + g_f^\dagger ) \\
\end{array}
\end{equation}
With these choices, $[H_f,H_v]=0$ for all faces $f$ and vertices
$v$.  For the two edges incident on a given vertex will be in
the boundary of some face, and after correcting for orientation
conventions this commutativity check reduces to
$[X \otimes X, Z \otimes Z^{-1}]=0$.  (See Figure \ref{fig:1}.)
Hence, for some constant $h>0$, we might define
$H_{\mbox{\footnotesize KE}} = 
h \sum_{f \in \mathcal{F}} H_f$.  Due to commutativity, the
kinetic energy Hamiltonian respects the groundstate degeneracy of
$H_\partial$.  Label $H=H_\partial + H_{\mbox{\footnotesize KE}}$.
We next show that the dimension of the groundstate degeneracy of 
total Hamiltonian 
\begin{equation}
H=H_\partial + H_{\mbox{\footnotesize KE}}
\label{totHam}
\end{equation}
(over $\mathbb{C}$) corresponds to the number of elements of
$H_1(\Gamma,\mathbb{F}_d)$.

\subsection{Homology class groundstates}
\label{subsec:homclass}

The goal of this section is to associate the degeneracy (dimension)
of this groundstate of $H=H_\partial + H_{\mbox{\footnotesize KE}}$
 to $\# H_1(\Gamma,\mathbb{F}_d)$.  We accomplish
this in two distinct cases for the manifold underlying $\Gamma$:
\begin{enumerate}
\item  
\label{it:closed}
The manifold is orientable, compact, and has no boundary, so
that $H_2(\Gamma,\mathbb{F}_d)=\mathbb{F}_d$.
\item  
\label{it:no_H2}
The manifold is compact with boundary 
and has $H_2(\Gamma,\mathbb{F}_d)=0$.  For homology is
a homotopy invariant, and such a surface retracts into its
one skeleton.
\end{enumerate}

\noindent{\bf Assertion:}
Let $\mathcal{H}_{\mbox{\footnotesize loop}}$ denote the groundstate
of $H_\partial$ and  $\mathcal{H}_{[\omega]}=\oplus_{\eta \in [\omega]} 
\mathbb{C}\ket{\eta}$. 
\begin{equation}
\mathcal{H}_{\mbox{\footnotesize loop}} \ = \ 
\oplus_{\omega \in \mbox{\footnotesize ker }\partial} 
\mathbb{C} \ket{\omega} \ = \ 
\oplus_{[\omega] \in H_1(\Gamma,\mathbb{F}_d)} \; \mathcal{H}_{[\omega]}
\end{equation}
Throughout this section, let 
$\pi={\# G}^{-1} \sum_{g \in G} g$.  Suppose either
Case \ref{it:closed} or Case \ref{it:no_H2}.
Then for each
$[\omega]$, the restriction of $\pi$ to $\mathcal{H}_{[\omega]}$
is a rank one projector whose (nonzero) image is an element
of $\mbox{ker }(H_\partial + H_{\mbox{\footnotesize KE}}) = \mbox{ker }H$.

To verify this, suppose $\ket{\omega}$ is the computational basis state of
some cycle $\omega \in C_1$ (i.e. $\partial \omega=0$.)
Then we may also speak of $[\omega] \in H_1(\Gamma,\mathbb{F}_d)$,
$\ket{\omega}$ is in the groundstate of $H_\partial$.  Label
\begin{equation}
\ket{[\omega]} \ {\buildrel \mbox{\scriptsize def} \over =} \ 
\pi_f \ket{\omega} \ = \ (\# G)^{-1}
\sum_{g \in G} g \ket{\omega} .
\end{equation}
It suffices for the Assertion to show the following.
\begin{itemize}
\item  If $\omega_1$ and $\omega_2$ each lie in $[\omega]$, then
$\ket{[\omega_1]}$ and $\ket{[\omega_2]}$ differ by a global phase.
\item  If $\ket{\omega} \neq 0$, then $\ket{[\omega]}\neq 0$.
\end{itemize}
This suffices to see the restriction of $\pi$
is a rank one projector, since the first item demands the rank
$\leq 1$ and the second demands the rank $\geq 1$.

We begin with the first item, writing
$\omega_1 - \omega_2 = \eta \in \mbox{im } \partial_2$.
Since the underlying manifold of $\Gamma$ is orientable,
suppose for convenience that all faces $f$ have positive
orientation.  Then for $\eta=\sum_{f \in S(\eta)} f$ we put
$g_\eta=\prod_{f \in S(\eta)} g_f$, implying
$\ket{\omega_1}=g_\eta \ket{\omega_2}$.  Note that
$g_\eta \pi_f = \pi_f g_\eta = \pi_f$.  Thus 
$\ket{[\omega_1]}= \pi_f g_\eta \ket{\omega_2} = \pi_f \ket{\omega_2}
=\ket{[\omega_2]}$.

We next demonstrate that $\pi_f|_{\mathcal{H}_{[\omega]}}$ 
has $\mbox{rank} \geq 1$.
As discussed in \S \ref{sec:stabilizer}, it suffices to show that
the trace of this projection, when restricted to the subspace
$\mathcal{H}_{[\omega]}$ which it preserves, is nonzero, and that
immediately follows if $\xi^\ell I_{d^n} \in G_f$ demands $\xi=1$.
For all other elements of $\mathcal{P}(n,d)$ are traceless
when restricted to $\mathcal{H}_{\mbox{\footnotesize loop}}$,
since $g_\eta \ket{\omega} = \ket{\omega + \partial \eta}$.
Case \ref{it:closed} and Case \ref{it:no_H2} differ somewhat.
In each case, multiples of the identity
in $G_f$ are products $g_\eta$ for $[\eta] \in H_2(\Gamma, \mathbb{F}_d)$.
In Case \ref{it:closed}, besides the empty product of the $g_f$
we also produce multiples of $I_{d^n}$ as the full product
$\prod_{f \in \mathcal{F}} g_f^k$, $0 \leq k \leq d-1$.
This corresponds to $H_2(\Gamma,\mathbb{F}_d)=\mathbb{F}_d$.
Yet for these products $\xi=1$, as may be verified at an
individual edge.  In Case \ref{it:no_H2}, there is no nontrivial
product of the $g_f$ which produces
a multiple of the identity.  This is due to the retraction
demanding $H_2(\Gamma,\mathbb{F}_d)=0$, the second homology of the
one complex we may retract onto.  Colloquially,
taking a sum of all faces will force a boundary edge to be acted
on nontrivially by $g_f$ for the single face it bounds.  Thus in
Case \ref{it:no_H2} the only multiple of the identity is the trivial
product of the $g_f$, and $\xi=1$ tautologically.
In each case, $\pi_f |_{\mathcal{H}_{[\omega]}}$ is not traceless
and hence has rank at least one.  Given the last paragraph, the
rank is exactly one.

Retracing the argument above, we may compute the image under
$\pi$ of the code space of $G_v$ is
$\oplus_{[\omega] \in H_1(\Gamma,\mathbb{F}_d)} \mathbb{C} \ket{[\omega]}$, 
which is also the code space of $G$.  
Since $\pi_f$ is a rank-one projector when restricted
to each $\mathcal{H}_{[\omega]}$, we have the following.
\begin{equation}
\mbox{dim}_{\mathbb{C}}
(\mbox{ groundstate of } H) \ = \ \# H_1(\Gamma,\mathbb{F}_d)
\end{equation}

\subsection{Groundstates on a punctured disk}
In practice, constructing physical realizations of Hamiltonians
corresponding to two-complexes without boundary is daunting.  It
is possible to simply identify opposite qudits on the square
fundamental domain of $S^1 \times S^1=\mathbb{R}^2/\mathbb{Z}^2$,
but this would require some sort of nonlocal coupling on the
boundary in addition to the standard lattice coupling.  Given a
lattice Hamiltonian that arises from electromagnetic coupling,
one could speculate about some kind of apparatus (perhaps 
involving fiber-optic cabling \cite{Bose})
which allows for interactions between boundary qudits.

\begin{figure}
\begin{center}
\includegraphics[scale=0.22]{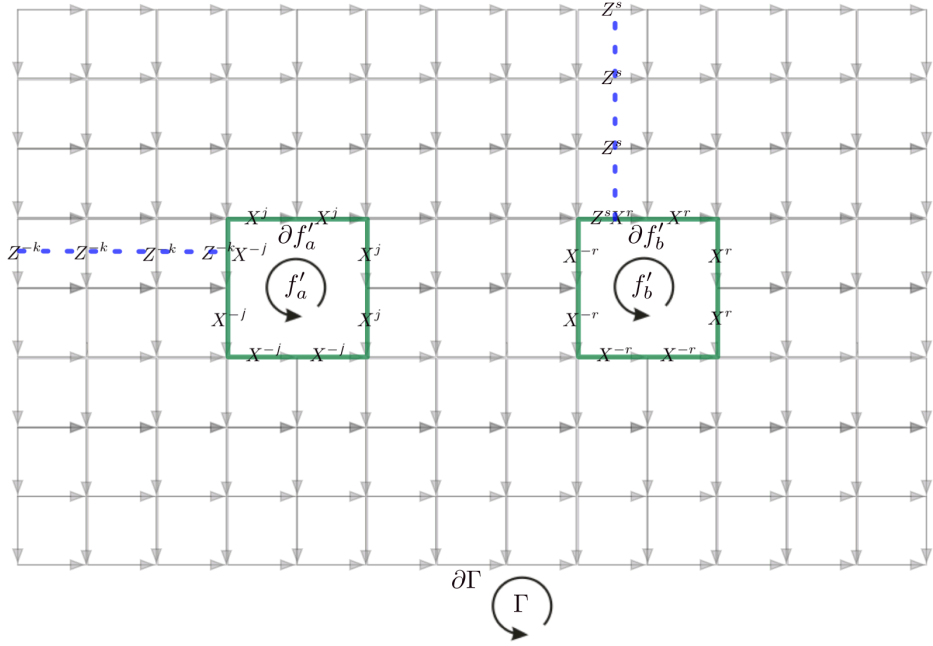}
\caption{\label{fig:6}  An oriented two complex $\Gamma$, which is a 
cellulation of a two punctured disk encoding two logical qudits in 
$n$ physical qudits.  Vertex operators $H_v$ are $k$ local where $k$ is the 
valence of the vertex whereas all face operators $H_f$ are $4$ local in this 
example.  Ground states are $+1$ eigenstates of the stabilizer group 
$G'$, but not all the stabilizer generators are independent. 
There are two independent non trivial cycles on $\Gamma$ which can be 
generated by closed loops of $X$ operators around the boundaries 
$\partial f'_a$ and $\partial f'_b$.  Similarly, there are two 
independent non 
trivial cycles on the dual $\tilde{\Gamma}$ which can be generated by 
strings of $Z$ operators that connect two independent pairs of boundaries 
of the complex.  Shown are the Pauli group operations $Z^kX^j$ on qudit 
$a$ and $Z^sX^r$ on qudit $b$.}
\end{center}
\end{figure}

Alternately, we might modify the homological groundstates to allow
for a surface with a boundary curve and punctures.  Consider a cellulation 
$\Gamma$
of a disk with $k$ punctures.  An example with $k=2$ is shown in 
Fig. \ref{fig:6}.
  Label the $j$-th puncture face $f'_j$ which has the same 
orientation as $\Gamma$.  Also label the outer boundary 
of the disk $\partial \Gamma$ and the boundaries of the 
$j$-th puncture $\partial f'_j$.
Analogous to the previous construction, the Hamiltonian on $\Gamma$ is defined
$H'=H_{\partial}+H'_{KE}$.  Here the kinetic term is modified so that the 
set of face operators does not include operators on the punctured faces 
$f'_j$, i.e.  $H'_{KE}=h\sum_{f\in\mathcal{F'}}H_f$, where 
$\mathcal{F'}=\mathcal{F}\setminus\{f\in\cup_{j=1}^k f'_j\}$.  Consequently, there are 
edges on the boundaries $\partial f'_j$ that are acted on by $X$ operators 
from faces on one side only.  Another way to see this is that all edges of 
the dual cellulation that cross the boundary $\partial f'_j$ share a 
common vertex located at $f'_j$ in $\Gamma$.
Each edge in $\Gamma$ has two vertices in $\mathcal{V}$, hence the 
product over all vertex operators is:
\begin{equation}
\prod_{v\in\mathcal{V}}g_v=I_{d^n}.
\label{zconst}
\end{equation}  
Not every edge in $\Gamma$ borders two faces in $\mathcal{F'}$, however, 
and the product over all face operators is: 
\begin{equation}
\prod_{f\in \mathcal{F'}}g_f=C_{\partial \Gamma}(X)\prod_{j=1}^k C_j(X),
\label{xconst}
\end{equation}
where $C_{\partial \Gamma}(X)=\prod_{e_j\in\partial \Gamma}X^{o_j}_{e_j}$ 
and $C_j(X)=\prod_{e_j\in\partial f'_j}X^{o_j}_{e_j}$.  The orientation 
$o_j=1$ if the edge $e_j$ is oriented in the same direction as the boundary 
on which the edge resides, and $e_j=d-1$ if the orientations are opposite.  

First we argue that the code space in nonempty.  Recall, the code states 
are defined as $+1$ eigenstates of the stabilizer group 
$G'=\langle \{g_f|f\in\mathcal{F'}\}\sqcup\{g_v\}\rangle$.  The operators    
$\langle \{g_v\}\rangle$ and $\langle\{g_f\}$ commute and the only additional 
relations obtained from the stabilizer group, embedded in Eqs. 
\ref{zconst},\ref{xconst}, guarantee that $(G'\bigcap \xi^jI_{d^n})=I_{d^n}$.  
We next show that the code space is 
$\mathcal{H}_{gr}=\mathcal{H}\mathcal(k,d)$ by considering the action of 
operators that commute with any member of $G'$ but act non trivially on 
$\mathcal{H}_{gr}$.  One such set of operators are non trivial 
$\mathbb{F}_d$ valued cycles on $\Gamma$ generated by $\{C_j(X)\}$.    We 
do not include the non trivial cycles generated by 
$C_{\partial \Gamma}(X)$ because by Eq. \ref{xconst} their action on the 
code subspace is not independent but can be generated by the cycles around 
the boundaries of the punctures.  A non trivial cycle on 
$\tilde{\Gamma}$ is generated by a string of $Z$ operations along a path 
$Path(j)$ that begins on an edge of $\partial f'_j$ and ends on an edge 
of $\partial \Gamma$ without touching other edges on puncture boundaries.  
We denote the generator of a such a cycle 
$C_j(Z)=\prod_{e_j\in Path(j)}Z^{o_j}_{e_j}$ where $o_k=1$ at the edge 
$e_k\in\partial f'_j$ if $e_k$ and $\partial f'_j$ share the same orientation 
and $o_k=d-1$ otherwise.  The other $o_j$ are chosen in a consistent way 
such that $[H',C(Z)_{f'_j}]=0$.  The operators on cycles  satisfy the 
commutation relations $C(Z)_{j}^aC(X)^b_{j}=\xi^{ab}C_j(X)^b C_j(Z)^a$ 
for $a,b\in\mathbb{F}_d$ as is easily verified by 
considering the action on the one intersecting edge $e\in\partial f'_j$.  
As such the set $R_j=\{C_j(Z)^aC_j(X)^b)\}_{a,b=0}^{d-1}$ generates 
a representation of the Pauli group $\mathcal{P}(1,d)$.  For sufficiently 
spaced punctures, all paths $Path(j)$ exist and the group 
$R=\langle\{R_j\}_{j=1}^k\rangle$ forms a representation of 
$\mathcal{P}(k,d)$.  
We then find that the ground subspace of $H'$ encodes $k$ qudits and the 
set $R_j$ performs local 
Pauli group operations on the $j$-th qudit.   

In a lattice implementation of our model Hamiltonian, the punctures may arise
as physical defects in the system.  Coding operations that correspond
to cycles around defects vividly illustrate 
the fact that even short ranged correlators (short relative to the
system size) in a topologically 
ordered state can have non-trivial values.

\section{Quantum Memory:  Input/Output and Error Detection}
\label{sec:memory}

We next describe how one might exploit abelian anyons
as quantum memories; the qubit case has been studied
thoroughly \cite{dennis}.  In the new setting of prime
level qudits, we must treat storage and retrieval of
quantum data.  It is also possible to generalize earlier
discussions of stabilizer operations on topologically stored
data while in code, but we will not treat that topic here.

\subsection{Storing qudits}

Placing quantum data into such a $\ket{[\omega]}$ is difficult.
For large lattices, this would be a special case of the
qudit state-synthesis problem.  Universal circuits of two-qudit
operators capable of reaching arbitrary $n$ qudit states are
known to scale exponentially with the number of qudits
\cite{bbo}.  In this section, we propose an alternative which
requires a number of stabilizer measurements that is linear in
the size of the lattice and also a sublinear number of entangling
gates.

For an orientable, connected, compact
surface of genus $g$, it is well known that
$H_1(\Gamma,\mathbb{F}_d)=(\mathbb{F}_d)^{2g}$.  (See e.g. \cite{topology}.)
We next describe how one might
transfer a qudit $\ket{\psi}$ stored within an ancilliary copy of
$\mathcal{H}(1,d)$ to the topologically ordered groundstate 
eigenspace of $H$, say $\mathcal{H}_{gr} \cong \mathbb{C}^{d^{2g}}$.

The suggestion for encoding is as follows.  We begin with
$\ket{\psi}=\sum_{j=0}^{d-1} \alpha_j \ket{j}$.  
Choose a copy of $\mathbb{F}_d \subseteq H_1(\Gamma,\mathbb{F}_d)$,
and let $[\omega]$ correspond to $1 \in \mathbb{F}_d$.
Choose $\omega \in [\omega]$, 
preferably with as few nonzero (excited) edges
as possible.  Now $j\omega$ is also a cycle for $0 \leq j \leq d-1$,
and by our choice $\{[j\omega]=j[\omega]\}_{j=0}^{d-1}$ contains 
distinct homology classes.  Using whatever unitaries are
convenient, we form
\begin{equation}
\ket{\tilde{\psi}} \ = \ \sum_{j=0}^{d-1} \alpha_j \ket{j\omega}
\end{equation}
For example, on a toric $\Gamma$ one might have $n$-sites and choose
a vertical or horizontal cycle on a square fundamental domain.  Then
the appropriate unitary would cost $O(\sqrt{n})$ gates.  Our
goal is to construct
$\ket{\psi_{\mbox{\footnotesize storage}}}=\sum_{j=0}^{d-1}
\alpha_j \ket{[j\omega]}$.  For the remainder of the construction,
note that all intermediate states are in the code space of the
stabilizer $G_v=\langle \{ g_v \} \rangle \subseteq \mathcal{P}(n,d)$.
Hence, we may correct for errors in this code at any time.  Also,
the scheme below might be thought of as arising from an error
correction to the stabilizer $G_f=\langle \{ g_f \} \rangle$.
Nonetheless, only 
$\ket{\psi_{\mbox{\footnotesize storage}}}$ is in the code space
of the full stabilizer $G$.  Arbitrary local errors are correctible
in the code space of $G$ since the normalizer of $G$ contains 
$\{Z_e,X_e \; ; \; e \in \mathcal{E}\}$
\cite{Gottesman,TA&M}.  Since this is clearly false for $G_v$,
one should perform the initialization above as quickly as possible.

We suppose an ordering of the faces $f \in \mathcal{F}$, say
$f_1$, $f_2$, \ldots, $f_L$, such that for each fixed $\ell$
the boundary of $f_\ell$ contains some edge $e_\ell$
which (i) is not within
the boundaries of $f_1$, $f_2$, \ldots, $f_{\ell-1}$
and (ii) does not intersect the support of $\omega$. 
This is not possible for the last face $f_L$, but we only
require this condition for $1 \leq \ell \leq L-1$.
To store the qudit beginning with $\ket{\tilde{\psi}}$,
we apply the following steps for each $f_\ell$.
\begin{itemize}
\item  Measure the eigenvalue of $g_{f_\ell}$, e.g. using an ancillary
qudit. (See \S \ref{sec:stab_measure}.)
The eigenvalue $\lambda$
will be an element of $\{ \xi^j \}_{j=0}^{d-1}$.  
\item  If $\lambda=1$, then 
the state has collapsed onto the
stabilizer $\langle \{g_v\} \sqcup \{g_{f_k} \; ; \; 1 \leq k \leq \ell \}
\rangle$ (by induction.)  Else, 
measuring $\xi^j$ accidentally performed the
collapse $P_j = (1/d) \sum_{k=0}^{d-1} \xi^{jk} g_{f_\ell}^k$,
which is in fact a projection \footnote{Why is this
a projection?  Consider the unitary $h=\xi^j g_{f_\ell}$ and
consider projection onto the stabilizer of $\langle \{h\} \rangle$.}
Let $e_\ell$ be the isolated edge as above.  
Since $Z_{e_\ell}^k g_{f_\ell} = \xi^k g_{f_\ell} Z_\ell^k$, we see that
$Z_{e_\ell}^{j} P_j=P_0 Z_{e_\ell}^{j}$.  Thus
an appropriate
power of $Z_{e_\ell}$ will fix the projection onto the unwanted
eigenvalue so that the final state lies within the the $+1$
eigenspace of $H_{f_\ell}$.
\end{itemize}

Applying the process of the last paragraph clearly produces
an element of $\mathcal{H}_{gr}$.
The applications of $H_f$, $P_j$,
and also $Z_{e}$ all respect $\mathcal{H}_{[j\omega]}$
for $0 \leq j \leq d-1$.  Note that
$\mathcal{H}_{gr} \cap \mathcal{H}_{[j\omega]} =
\mathbb{C} \ket{[j\omega]}$.  
If $S$ denotes the superoperator
of the above sequence of measurements and unitary maps,
then equivalently we have shown $S(\mathcal{H}_{\mbox{\footnotesize loop}})
\subseteq \mathcal{H}_{gr} \cap \mathcal{H}_{[j \omega]}$.
Equality is immediate after noting
$S \ket{[j \omega]}=\ket{[j \omega]}$.
 
However, the effect of the superoperator on relative phases
is still unclear.  Given the global
phase on $\ket{\omega}$, there is a natural global phase
on $\ket{[\omega]}=\pi \ket{\omega}$.
With the argument above, we have actually verified that
$S \ket{0} = \mbox{e}^{i \varphi_0} \ket{[0]}$,
$S \ket{\omega} = \mbox{e}^{i \varphi_1} \ket{[\omega]}$,
$S \ket{2\omega} = \mbox{e}^{i \varphi_2} \ket{[2\omega]}$, etc.
Thus perhaps $\ket{\psi_{\mbox{\footnotesize storage}}}=\sum_{j=0}^{d-1}
\mbox{e}^{i \varphi_j} \alpha_j \ket{[j\omega]}$.  We argue that all
of these relative phases are in fact equal.
For in terms of the observed eigenvalues,
\begin{equation}
S \ = \ \prod_{j=1}^{L-1} Z_{e_\ell}^{\pm j} P_j(f_\ell)  \ = \ 
\prod_{j=1}^{L-1} P_0(f_\ell) Z_{e_\ell}^{\pm j} \ = \ 
\pi \prod_{\ell=1}^{L-1}
Z_{e_\ell}^{\pm j}
\end{equation}
By choice of the support of
$\ket{\omega}$, also the support of $\ket{j\omega}$, we have
$\prod_{\ell=1}^{L-1} Z_{e_\ell}^{\pm j} \ket{j \omega}=1$.
Thus, the applying the superoperator $S$ to $\ket{\tilde{\psi}}$ produces
$\ket{\psi_{\mbox{\footnotesize storage}}}=\sum_{j=0}^{d-1}
\alpha_j \ket{[j\omega]}$, given that we may choose the
$\{e_\ell\}_{\ell=1}^{L-1}$ to be disjoint from the support of
$\omega$.

\subsection{Retrieval}

Thus we next consider retrieval of a qudit stored as in the
last subsection, i.e. swapping the data in a topological qudit
with that encoded in some ancilla qudit.  Physically, this is
more intricate than encoding, which amounts to creating a
cycle class $\ket{\omega}$ and then applying stabilizer corrections
for $\{g_f\}_{f \in \mathcal{F}}$ generating $G_f$.

For retrieval, the central point is that we may apply a logical
$X$ operation to the encoded qudit using $O(\sqrt{n})$ gates.
To see this, for $\omega=\sum_{e} n_e e$ let
$X^{\otimes \omega}=\otimes_{e \in \mathcal{E}} X_e^{n_e}$.
This might be thought of as a creation operator of an excitation
of the loop $\omega$, and moreover
$X^{\otimes \omega}$ is an element of the centralizer of $G$ not
contained within $G$.  As such, it preserves the code space,
and one readily verifies that it must map
$\ket{[j\omega]} \mapsto \ket{[(j+1)\omega]}$, up to global phase.
Hence, we may apply controlled-$X$ operations targetting the
topological qudit using $O(\sqrt{n})$ physical controlled-$X$
operations.

We next consider a controlled-$X$ operation controlled on the topological
qudit and targetting an ancilla.  One must choose a cycle in the
dual complex to $\Gamma$ according to $\omega$, say $\eta$.
For example, an earlier work \cite[Fig.3]{dennis} depicts a picket
fence dual to a loop generator of the first homology group of a torus.
In order to perform the required controlled-$X$, follow these steps.
\begin{itemize}
\item  Prepare a second ancilla.
Then prepare this second ancilla so that the $Z$ eigenstate of
the ancilla measures $Z^{\otimes \eta}$.
\item  Perform the controlled-$X$ contingent on this second ancilla.
\item  Disentangle, i.e. reverse the qudit gates of the first step.
\end{itemize}
Consquently, we can perform either controlled-$X$ to or from the
topologically encoded qudit.

The ability to perform a two-qudit controlled-$X$ gate implies
the ability to perform controlled modular addition.  The composition
begins with a single controlled increment triggering when the control
carries $\ket{1}$, continues with two controlled increments
when the control carries $\ket{2}$, etc.  The entire circuit thus
realizes a controlled modular addition in a number of controlled-$X$
gates roughly the triangular number of $d$.  Controlled modular
subtraction is similar.

Finally, modular addition and subtraction allow us to SWAP the topological
qudit to an ancilla.  For bits, the standard three CNOT swap
relies on the fact that CNOT exclusive-or's one bit to another.
Thus the CNOTs perform $b_1 b_2 \mapsto b_1(b_1\oplus b_2) \mapsto
b_2(b_1\oplus b_2) \mapsto b_2 b_1$.  In like manner, we may
perform suitably controlled and targetted additions and subtractions
for the following sequence of dit operations:
\begin{equation}
d_1 d_2 \ \mapsto \ d_1 (d_1 + d_2) \ \mapsto \ 
(-d_2) (d_1+d_2) \ \mapsto \ (-d_2) d_1
\end{equation}
Hence, modifying gates so that a control symbol with a $+$ or $-$ target
means to add or subtract the control respectively, we have the following
diagram:
\[
\Qcircuit @C=1em @R=2em
{
& \qswap \qw & \qw & & & & \qw & \ctrl{1} & \gate{-} & \ctrl{1} & \gate{-1} &
\qw & \\
& \qswap \qwx \qw & \qw & & \ustick{\cong} & & \qw & \gate{+} & \ctrl{-1} & 
\gate{+} & \qw & \qw
}
\]
We have not described how to complete the
gate $\ket{j} \mapsto \ket{d-j}$ on the topologically ordered state.
Rather than do so, we claim the top line as the ancilla.
This is also improves the cost of the controlled additions.

\subsection{Modified constructions using ancillary qudits}
\label{secancstab}

In the quantum circuit model of computation ancillary particles are often used
as a means to assist in gate operations and as an 
entropy dump during error correction cycles.
In the context of surface codes it is tempting 
to borrow this idea and place qudits at the center
of each face and on each vertex of the cellulation $\Gamma$, so that the appropriate
stabilizer checks might be done in place (see Figure \ref{fig:1}b).
Recall, any state may be projected
into the groundstate of the topologically ordered
Hamiltonian $H=H_\partial + H_{\mbox{\footnotesize KE}}$
using stabilizer checks to the Pauli tensors
$\{g_v\}\sqcup \{g_f\} \subset \mathcal{P}(n,d)$ 
(\S \ref{subsec:homclass}.)
Each individual stabilizer check may then be
performed using a certain sequence of two-qudit gates
and a neighboring ancilla (\S \ref{sec:stab_measure}.)  
In fact, this basic observation presents an auxilliary Hamiltonian
which also computes the same topological order as the
original.  Namely, on the face-edge-vertex qudit system,
one may build a Hamiltonian which is in the groundstate
iff all the stabilizer checks $g_v$ and $g_f$ are satisfied.
For $g_v$, suppose we use $\Sigma_e^v$ for the sum gate
targetting the qudit of vertex $v$ and take
${\bf n}_v$ to be the qudit number operator on $v$.  Then
\begin{equation}
\tilde{H}_v \ = \ 
\prod_{[\ast,v]=e} \Sigma_e^v \prod_{[v,\ast]=e} (\Sigma_e^v)^{-1} 
{\bf n}_v
\prod_{[\ast,v]=e} (\Sigma_e^v)^{-1} \prod_{[v,\ast]=e} \Sigma_e^v 
\end{equation}
Then $\ket{\psi}$ is in the groundstate of $\tilde{H}_v$ iff 
$g_v \ket{\psi}=\ket{\psi}$.  Similarly, fix a face
$f \in \mathcal{F}$ with $\partial f = \sum_{j=1}^{\ell} n_j e_j$
for $n_j \in \{1,d-1\}$.  We take ${\bf F_f}=\prod_{j=1}^{\ell}
(\mathcal{F}_d)_{e_j}$ and $U_f = \prod_{j=1}^{\ell}
(\Sigma_{e_j}^f)^{n_j}$ for $\Sigma_e^f$ the sum gate targetting
the $f$ qudit.  Then for ${\bf n}_f$ the number operator of the
qudit at the center of the face $f$, we label
\begin{equation}
\tilde{H}_f \ = \ {\bf F_f} U_f {\bf F_f}^\dagger \; {\bf n}_f \;
{\bf F_f} U_f^\dagger {\bf F_f}^\dagger
\end{equation}
As before, we see that $\ket{\psi}$ is in the groundstate of
$\tilde{H}_f$ iff $g_f \ket{\psi}=\ket{\psi}$.
Thus for $h > 0$ and $U>0$, if $\tilde{H}=U\sum_v \tilde{H}_v +
h\sum_f \tilde{H}_f$, then the groundstate of $\tilde{H}$ is also
the code space of $G=\langle \{g_v\}\sqcup \{g_f\} \rangle$,
i.e. the topologically ordered groundstate spanned by
\hbox{
$\{ \ket{[\omega]} \; ; \; [\omega] \in H_1(\Gamma,\mathbb{F}_d)\}$.}

We finish this section describing another utility 
for ancillary particles, namely to mediate
many body interactions present in the Hamiltonian $H$ (Eq. \ref{totHam})
 using more physically motivated binary interactions.  
Consider the vertex constraint term
$H_v=-(g_v+g_v^{\dagger})$ where the valence at that vertex is $k$.  
This $k$-local interaction can be obtained as a perturbative limit
of $2$-local interactions between each $d$-level qudit incident at $v$ and a $k$-level
ancillary qudit $a$ located at the vertex.  Begin with a local Hamiltonian for the ancilla
 $H_a=-E_a \ket{0}_a\bra{0}$, 
and a perturbing interaction $V_a=J_v \sum_{r=1}^k(Z^{o_r}_{e_r}\otimes \ket{r-1}\bra{r}+h.c.)$, where $E_a\gg |J_v|$ and the edge orientations give $o_j=1$ if $e_j=[*,v]$ and $o_j=d-1$ if $e_j=[v,*]$.  By construction, the lowest nontrivial, i.e. non identity, contribution to coupling in the ground subspace 
$\mathcal{H}_{gr}=\ket{0}_a\bra{0}(H_0+V)\ket{0}_a\bra{0}$ is the effective Hamiltonian
$H_{v\rm eff}=U(H_v+O(\epsilon))$ where $U=(-1)^kE_a(J_v/E_a)^k$ with an error term of norm 
$||\epsilon||\ll1$.  By judicious choice of ${\rm sign}(J_v)$ it is possible to fix $U>0$.  A similar argument applies to building 
the face constraint $H_f$ using a $j$-level ancilla $b$ located at face $f$
to mediate interactions between all $j$ edges 
on the boundary of $f$.  Here we choose $H_b=-E_b \ket{0}_b\bra{0}$ and $V_b=J_f \sum_{r=1}^j(X^{o_r}_{e_r}\otimes \ket{r-1}\bra{r}+h.c.)$  such that $H_{f\rm eff}=h(H_f+O(\epsilon))$, where $h=(-1)^j E_b(J_f/E_b)^j$.  
These mediator qudits could be placed on all the vertexes and faces of $\Gamma$ to build an effective Hamiltonian in the subspace spanned by states with all ancillae in the $\ket{0}$ state.

An argument in Ref. \cite{Oliveira} suggests that an effective Hamiltonian between spins on a two complex can be built using such mediating interactions that closely approximates a target Hamiltonian projected to its ground subspace $H_{gr}=P_{gr}HP_{gr}$.  In the present context this would imply that for sufficiently large energies $E_a,E_b$ both the degeneracy of the ground subspace of $H=H_\partial + H_{\mbox{\footnotesize KE}}$ as well as the energy gap to the excited states could be accurately approximated by a model built from a sum of effective vertex and face operators.  An analysis regarding the validity of such constructions for topologically ordered states is wanting, but is outside the scope of this work.

\section{Other Homological Groundstates}
\label{sec:extensions}

We have originally presented the case of groundstates
for $H_1(\Gamma,\mathbb{F}_d)$ for $d$ prime, in order
to present the new orientation conventions in the simplest
possible context.  
This section describes a construction for homological order
on dits whose number of levels is not prime but rather a prime
power.  Homological order for arbitrary composite $d$ follows
immediately through a tensor product of the prime-power Hamiltonians.

\subsection{Homology $\mathbb{F}_{d^\ell}$ stabilizer codes}

The hypothesis in the main text has been that qudits have $d$
levels, for $d$ a prime so that each $\ket{j}$ is associated
to an element of $\mathbb{F}_d$.  Recent work \cite{TA&M} 
extends stabilizer techniques to the finite
fields of order $d^\ell$, i.e. $\mathbb{F}_{d^\ell}$, which
exist for any $\ell \geq 1$.  

The generic $\mathbb{F}_{d^\ell}$
constitute all fields $\mathbb{F}$ with $\# \mathbb{F} < \infty$,
so this is (perhaps) the most general field for which a stabilizer code
makes sense.  The most typical construction of
$\mathbb{F}_{d^\ell}$ is to consider the polynomial ring
$\mathbb{F}_d[x]$ and divide out relations in the ideal generated
by some irreducible polynomial $f(x)=x^\ell + a_{\ell-1}x^{\ell-1} + \ldots
+ a_0$, $a_j \in \mathbb{F}_d$.  It is typical to label
$\alpha \in \mathbb{F}_{d^\ell}$ as the adjoined root corresponding
to the class of $x$.  The Galois group of the extension
$\mathbb{F}_{d^\ell}$ over $\mathbb{F}_d$, say $K$, then acts as permutations
of the roots of $f(x)$.  Note that $\mathbb{F}_{d^\ell}$ is
a vector space over the scalars $\mathbb{F}_d$.  Moreover,
multiplication by any fixed $a \in \mathbb{F}_{d^\ell}$ may be
viewed as a $\mathbb{F}_d$-linear map, with an associated matrix
with entries in $\mathbb{F}_d$.  Computing the trace of this matrix
creates a map 
\hbox{$\mbox{Trace}_{\mathbb{F}_{d^\ell}/\mathbb{F}_d}: 
\mathbb{F}_{d^\ell} \rightarrow \mathbb{F}_d$}.  Another characterization
is that $\mbox{Trace}_{\mathbb{F}_{d^\ell}/\mathbb{F}_d}(x)
= \sum_{\kappa \in K} (\kappa \cdot x)$.
To ground the discussion, let us review not extensions over
finite fields but rather $\mbox{Trace}_{\mathbb{C}/\mathbb{R}}(z)=
z + \overline{z}=2\mbox{Re}(z)$.  The complex conjugate is the Galois
action that interchanges $i \leftrightarrow -i$, for
$\mathbb{C}=\mathbb{R}[x]/(x^2+1)$.  We might instead
form a $2 \times 2$ matrix for multiplication by $z=x+iy$, which results in
$\mu_z=x \ket{0}\bra{0} -y\ket{1}\bra{0}-y\ket{0}\bra{1}+x\ket{1}\bra{1}$
with trace $2x$.

For $\mathbb{F}_{d^\ell}$ extending $\mathbb{F}_d$
the Galois group $K$ is cyclic of order $\ell$, generated
by $x {\buildrel {\tiny \kappa} \over \mapsto } 
x^d$ for $x \in \mathbb{F}_{d^\ell}$.  Now $\kappa$ generates
a one-qu$d^\ell$it unitary $U_\kappa$ by
$U_\kappa\ket{ x } = \ket{ \kappa \cdot x}$, and
the corresponding diagonal unitary on the entire lattice
will be denoted $\tilde{U}_\kappa$.

\subsubsection{Fourier transforms for $\mathbb{F}_{d^\ell}$}

Having reviewed the machinery of finite fields, we next review
what one would mean by a stabilizer code of Pauli matrices indexed
by it \cite{TA&M}.  Since our earlier qudit operators $X$ and $Z$
for $\mathbb{F}_d$ had order $d$, we might instead claim to have
constructed an $X$ operator and a $Z$ operator for each $a \in \mathbb{F}_d$,
i.e. $X^a$ and $Z^b$.  For $\mathbb{F}_{d^\ell}$, we do not take
operator powers.  Label 
$\mathcal{H}(1,d^\ell)=\oplus_{a \in \mathbb{F}_{d^\ell}} 
\mathbb{C}\{\ket{a}\}$.  Then suitable definitions are as follows,
where we define $\xi=\mbox{exp}(2\pi i/d)$.
\begin{equation}
\left\{
\begin{array}{lcl}
X(a) \ket{b} & = & \ket{a+b} \\
Z(a) \ket{b} & = & \xi^
{\mbox{\footnotesize Trace}_{\mathbb{F}_{d^\ell}/\mathbb{F}_d}(ab)}
\ket{b} \\
\end{array}
\right.
\end{equation}
For $\ell=1$, this generalizes the powers of earlier Pauli operators.
Furthermore, with these conventions we have a commutator relation:
\begin{equation}
X(a) Z(b) \ = \ \xi^
{\mbox{\footnotesize Trace}_{\mathbb{F}_{d^\ell}/\mathbb{F}_d}(ab)}
Z(b) X(a)
\end{equation}
Finally, let $\mathcal{H}(n,d^\ell)=\mathcal{H}(1,d^\ell)^{\otimes n}$.
In a slight abuse of notation, for
$a,b \in (\mathbb{F}_{d^\ell})^n$ we will write
$a \bullet b = \mbox{Trace}_{\mathbb{F}_{d^\ell}/\mathbb{F}_d}
(a_0 b_0 + a_1 b_1 + \cdots + a_{n-1}b_{n-1})$.  Then we may
generalize the earlier commutatator formula for Pauli tensors as 
\begin{equation}
\begin{array}{lc}
[ Z(b_0) \otimes Z(b_1) \otimes \cdots \otimes Z(b_{n-1}) ] \\
\ [ X(a_0)\otimes X(a_1) \otimes \cdots \otimes X(a_{n-1}) ] & = \\
\xi^{a \bullet b}[ X(a_0)\otimes X(a_1) \otimes \cdots \otimes X(a_{n-1}) ] \\
\ [ Z(b_0) \otimes Z(b_1) \otimes \cdots \otimes Z(b_{n-1}) ]
\end{array}
\end{equation}
Given this relation, one may define $\mathcal{P}(n,d^\ell)$ to be that
group generated by products of Pauli tensors indexed by
$\mathbb{F}_{d^\ell}$, as above.  Continuing, we may consider sets of
particular Pauli tensors $X(a_0)Z(b_0)\otimes \cdots \otimes
X(a_{n-1})Z(b_{n-1})$ and consider the stabilizer subspaces
of the subgroup $G \subset \mathcal{P}(n,d^\ell)$ they generate.
The error lengths of such code are studied in detail \cite{TA&M}.

Before considering which of these stabilizer codes arise
as topological orders, we add a point omitted in the original
treatments.  Namely, we wish to propose quantum circuits for
the appropriate stabilizer checks.  We suppose the existence
of a number operator measurement which can
output classical values in the finite field, say abusively
$n=\sum_{a \in \mathbb{F}_{d^\ell}} a \ket{a}\bra{a}$.
Then as with $\mathbb{F}_d$, stabilizer checks
would follow given an appropriate Fourier transform
$\mathcal{F}_{d^\ell}: \mathcal{H}(n,d^\ell) \rightarrow
\mathcal{H}(n,d^\ell)$ which
maps $X(a)$ eigenstates to $\ket{a}$.  This
leads one to guess we should define
$\mathcal{F}_{d^\ell} \ket{a} = (d^\ell)^{-1/2}\sum_{b \in \mathbb{F}_{d^\ell}}
Z(a) \ket{b}$, i.e.
\begin{equation}
\mathcal{F}_{d^\ell} \ {\buildrel \mbox{\tiny def} \over =} \ 
(d^\ell)^{-1/2} \sum_{a,b \in \mathbb{F}_{d^\ell}}
\xi^{\mbox{\footnotesize Trace}_{\mathbb{F}_{d^\ell}/\mathbb{F}_d}(ab)} 
\ket{b}\bra{a}
\end{equation}
However, note that $X(a)$ now has degenerate eigenspaces when
$\ell \geq 2$.  Thus, it is not clear the the above equation
actually defines a unitary matrix.

We briefly comment on why unitarity holds.  For convenience,
let us drop the subscript from the appropriate trace maps.
A computation reveals that
the unitarity assertion is equivalent to knowing that for any
fixed $a \in \mathbb{F}_{d^\ell}$ which is nonzero,
\begin{equation}
\sum_{b \in \mathbb{F}_{d^\ell}} \xi^{\mbox{\footnotesize Trace}(ab)} \ 
{\buildrel {\mbox{\tiny ?}} \over =} 
\ 0
\end{equation}
Since $a \neq 0$ has a multiplicative inverse, this amounts to
\begin{equation}
\label{eq:is_zero?}
\sum_{b \in \mathbb{F}_{d^\ell}} \xi^{\mbox{\footnotesize Trace}(b)} \ 
{\buildrel {\mbox{\tiny ?}} \over =} \ 0
\end{equation}
Now suppose we use $\alpha$ to denote the formally adjoined
root of $f(x)$ in $\mathbb{F}_{d^\ell}=\mathbb{F}_d[x]/(f(x))$.
Then since every equivalence class may be written as a polynomial
of degree less than $\ell$, we see that
$\{\alpha^j\}_{j=0}^{\ell-1}$ is a basis of $\mathbb{F}_{d^\ell}$
over $\mathbb{F}_d$.  In terms of the last basis, we might
express a generic polynomial class in coordinates as
$b=b_{\ell-1}x^{\ell-1}+b_{\ell-2}x^{\ell-2}+\cdots + b_0$ for
$b_j \in \mathbb{F}_d$.  Then Equation \ref{eq:is_zero?} becomes
\begin{equation}
\begin{array}{l}
\sum_{b_{\ell-1}=0}^{d-1}
\sum_{b_{\ell-2}=0}^{d-1} \cdots
\sum_{b_0=0}^{d-1} \\ 
\quad [ \xi^{\mbox{\footnotesize Trace}(\alpha^{\ell-1})}]^{b_{\ell-1}}
[ \xi^{\mbox{\footnotesize Trace}(\alpha^{\ell-2})}]^{b_{\ell-1}}
\cdots
[ \xi^{\mbox{\footnotesize Trace}(1)}]^{b_0}
\\ \ {\buildrel {\mbox{\tiny ?}} \over =} \ 0 \\
\end{array}
\end{equation}
This will in fact be zero, unless all
$\mbox{Trace}(\alpha^j)=0\mbox{ mod }p$, $0 \leq j \leq \ell-1$.
A standard construction in field
extensions is to form the discriminant of a basis, for our
basis $\Delta=\sum_{j,k=0}^{\ell-1} \mbox{Trace}(\alpha^{j+k}) \ket{k}\bra{j}$.
For a given basis, it is not possible that this matrix $\Delta$
have determinant zero in $\mathbb{F}_d$
\cite[Thm2.37,pg.61]{field_theory_ref}.  
Since the
first column of $\Delta$ can not then be zero, 
all $\mbox{Trace}(\alpha^j)$ may not be zero,
and unitarity of $\mathcal{F}_{d^\ell}$ follows.

\subsubsection{Homological order for $\mathbb{F}_{d^\ell}$}

The chain complex for computing $H_1(\Gamma,\mathbb{F}_{d^\ell})$
extends our early discussion by allowing for coefficients of
the vertices, edges, and face to be within $\mathbb{F}_{d^\ell}$,
which in context is $\ell$ copies of $\mathbb{F}_d$ since only
the additive structure is relevant.  Yet the previous section
has nontrivally extended our definition of $X$ and $Z$ operators
to account for field multiplication, and these operators may
be used to form a homological order on the physical system
in which qu$d^\ell$its (with $d^\ell$ levels) are associated 
to the edges of $\Gamma$:
\begin{itemize}
\item  For each vertex, we may again set
$g_v = \prod_{e=[\ast,v]} Z_e(1) \prod_{e=[v,\ast]} Z_e(-1)$ and
$H_v  = -(g_v + g_v^\dagger)$.  Then again $H_\partial = U
\sum_{v \in \mathcal{V}} H_v$.
\item  Again set
$g_f = X_{e_1}(o_1) X_{e_2}(o_2) X_{e_3}(o_3) \ldots
X_{e_p}(o_{p})$, where $\partial f = \sum_{\ell=1}^{p} o_\ell e_\ell$.  Put
$H_f = -(g_f + g_f^\dagger )$.  Given the generalization
of the commutators of the new $X$ and $Z$ operators,
$[H_f,H_v]=0$ for any $f$,$v$.  Then for $h> 0$, 
$H_{\mbox{\footnotesize KE}} = h \sum_{f \in \mathcal{F}} H_f$. 
\item  So $H=H_\partial + H_{\mbox{\footnotesize KE}}$.
A similar argument to that given before produces a basis
$\ket{[\omega]}$ of the groundspace of $H$, as
$[\omega]$ runs over all elements of $H_1(\Gamma,\mathbb{F}_{d^\ell})$.
\item  These groundstates may again be viewed as a stabilizer code
of $G=\langle \{ g_v,  g_f\}\rangle 
\subsetneq \mathcal{P}(n,d^\ell)$.  
Stabilizer checks can be performed as before (see \S\ref{sec:stab_measure}).
The only required modifications are that the quantum circuit uses
the new Fourier transform over $\mathbb{F}_{d^\ell}$
to measure $X(a)$ operators and powers thereof and the number operator
measurement now takes values in $\mathbb{F}_{d^\ell}$.
\end{itemize}
We close with one further comment.  Recall $\tilde{U}_\kappa$ 
which act on each qud$^\ell$it as
$U_\kappa \ket{a}=\ket{\kappa \cdot a}$ for $\kappa$
the generator of the cyclic Galois group of $\mathbb{F}_{d^\ell}$
extending $\mathbb{F}_d$.
Now $\tilde{U}_\kappa H = H \tilde{U}_\kappa$, as one can verify
directly using $H_f$ and $H_g$.  Thus we may view $\tilde{U}_\kappa$
or more generally the Galois action as a symmetry of the
topologically ordered groundstate.  Also,
$\pi_\kappa=\ell^{-1} \sum_{j=0}^{\ell-1} U_\kappa^j$ will then
act as a projection collapsing the groundstate
associated to elements of $H_1(\Gamma,\mathbb{F}_{d^\ell})$ onto
the groundstate parametrized by $H_1(\Gamma,\mathbb{F}_d)$
as constructed in \S \ref{sec:prime}.  In terms of Hamiltonians,
$\pi_\kappa$ projects onto the groundstate of
$H_\kappa=-(U_\kappa+U_\kappa^\dagger)$, whose physical
significance is unclear.

\section{$\mathbb{Z}/d \mathbb{Z}$ Gauge Theory and Anyonic Excitations} 

\label{anyons}
In our treatment of code subspaces, we 
have used the isomorphism between 
spins on a surface and one-chains on a two complex
to label the ground states of the Hamiltonian $H$
in terms of homology equivalence classes.  
The language of cell complexes also carries over to describe 
the excited states.  If we 
identify the ground subspace of $H$ as the vacuum
then excited states are labeled by $\mathbb{F}_d$ valued boundaries
of one chains on the complex $\Gamma$ or the dual complex $\tilde{\Gamma}$. 
These excitations can be viewed as massive particles with definite 
statistics.

In this section we show by construction that our model is a 
$\mathbb{Z}/d\mathbb{Z}$
gauge theory with quasi-particles corresponding to dyonic combinations of 
charge and flux.  
These quasi-particles have 
abelian anyonic statistics.  We provide an algorithm in terms of an 
interferometer circuit for measuring components of the scattering 
matrix.

\subsection{Stabilizer errors as abelian anyons}
\label{sec:particles}

Consider a two-complex $\Gamma$ with a physical system of
qudits associated to each edge and a topologically ordered
Hamiltonian $H$ as above.  We have already seen how to
associate a basis of the groundstate eigenspace with elements
of $H_1(\Gamma,\mathbb{F}_d)$.  As stabilizer states, it is
well known that the groundstates are entangled.  
Abelian anyons arise as entangled
excitations of this system.  In the qubit case,
such excitations always arise in pairs \cite{Kitaev}.
In our generalization, this is also true, and an excitation
$\ket{j}$ is always paired to an excitation $\ket{d-j}$.

The linear algebra for constructing a charge anyon is as follows.
First, choose two vertices $v_1$ and $v_2$
of $\Gamma$ on which the anyon should
reside with charges $j$ and $d-j$ respectively.
Choose a chain $\omega$ with $\partial \omega = jv_1+(d-j)v_2$.
Recall from \S \ref{subsec:homclass} the projection
$\pi = (\# \mathcal{F})^{-1} \sum_{f \in \mathcal{F}} g_f$
which projects onto the stabilizer code of all the face
operators $g_f=\otimes_{e \in \partial f}(X_e)^\pm$.
We set
\begin{equation}
\ket{\psi_{\mbox{\footnotesize charge anyon}}} \ = \ \pi \ket{\omega}
\end{equation}
The resulting state
is an excited state of $H_\partial$ whose eigenenergy
is $4U(1-\mbox{Re}(\mbox{e}^{2i\pi j/d}))$ above ground.  
It is \emph{not} independent of the choice of $\omega$,
and this in fact allows for an interesting geometric
interpretation of the error length of the associated
stabilizer code \cite{dennis}.

For let $\omega_1$ and $\omega_2$ be two such choices,
with $\ket{\psi_1}$ and $\ket{\psi_2}$ the resulting
anyon states.  Then $\omega_1 - \omega_2$ is a cycle, and
\begin{equation}
\ket{\phi}\ 
{\buildrel {\mbox{\tiny def}} \over = } \ 
\pi(\ket{\omega_1}-\ket{\omega_2}) \ = \ 
\ket{\psi_1}-\ket{\psi_2}
\end{equation}
is the ground state eigenket associated to
$[\omega_1 - \omega_2] \in H_1(\Gamma,\mathbb{F}_d)$.
Hence, if we encounter such a charge anyon excitation
which has sullied a qudit encoded in the groundstate of $H$,
then correcting it amounts to choosing an cancelling anyon
or equivalently to choosing
a cycle on $\Gamma$.  If the dual charges of
the anyon are separated by roughly half the diameter of
the two-complex, then this choice
is likely to cause an error.  Yet for nearby dual charges
one might reasonably guess $[\omega_1 -\omega_2]=[0]$.
In particular, if $\Gamma$ were to cellulate the square
fundamental domain of a torus using $n$ qudits
on the edges (implying $\Theta(\sqrt{n})$ qudits on a side,) then
we would expect an error length for the associated stabilizer
code to be roughly $O(\sqrt{n})$ \cite{dennis}.

Similar comments apply not only to charge anyons
but also flux anyons \cite{Kitaev}.  Here, one chooses
a path in the dual complex to $\Gamma$, i.e. a sequence of
connected faces.  Let $\ket{[0]}$ be the homological groundstate
associated to $[0] \in H_1(\Gamma,\mathbb{F}_d)$.
A flux charge of multiplicity $j$ on the
endpoints of the face path is associated to
\begin{equation}
\ket{\psi_{\mbox{\footnotesize flux anyon}}} \ = \ \pi_v 
g_{f_1}^{\pm j} g_{f_2}^{\pm j} \cdots g_{f_\ell}^{\pm j} \ket{[0]}
\end{equation}
where $\pi_v = (\# \mathcal{V})^{-1} \sum_{v \in \mathcal{V}} g_v$
and the path consists of faces $f_1,f_2,\ldots,f_\ell$
with the signs allowing for orientation.  The flux anyon theory
follows quickly by considering the charge anyons of the dual
two-complex to $\Gamma$, say $\tilde{\Gamma}$.  Faces of $\Gamma$
become vertices of $\tilde{\Gamma}$ while vertices become
faces, and the graph of $\tilde{\Gamma}$ arises by connecting
vertices corresponding to incident faces of $\Gamma$.
Suitable hypotheses on the cellulation of the underlying
two-manifold of $\Gamma$ will cause this dualization procedure
to be well behaved \cite{topology}, and one might associate
charge-anyonic observation of flux anyons and vice versa with
pairings exploited in the proof of Poincar{\'e} duality.

\subsection{Quasi-particle statistics}

We next wish to study such anyon states, i.e. errors of the
stabilizer code as above.  New notation for the excitations follows.
A charge $a\in \mathbb{Z}/d\mathbb{Z}$ at 
vertex $v$ is labeled by the state $\ket{(a,0;(v,-))}$ such that 
$\bra{(a,0;(v,-))}g_v\ket{(a,0;(v,-))}=\xi^{a}$.
Similarly, flux \
$b\in\mathbb{Z}/d \mathbb{Z}$ at face $f$ is labeled by the state 
$\ket{(0,b;(-,f))}$ such that
$\bra{(0,b;(-,f))}g_f^{\dagger}\ket{(0,b;(-,f))}=\xi^{b}$.  
A dyon refers to a bound state 
of charge and flux at vertex $v$ and face $f$ neighboring each other,
i.e. 
$[v,\ast]\in\partial f$ or $[\ast,v]\in\partial f$ and
$(a,b)\in (\mathbb{Z}/d\mathbb{Z})^2$.  The state of such a dyon
in Hilbert space will be denoted
$\ket{(a,b;(v,f))}$.  For simplicity we restrict our discussion to simply connected
compact surfaces with boundary such that the ground (vacuum) state is 
nondegenerate \footnote{In general $\ket{(a,b);(v,f)}$ describes an equivalence 
class of pure states which results from applying $X^a_e Z^{-b}_e$ to 
\emph{any} groundstate.  For a degenerate vacuum, particle creation,
followed by braiding and annihilation can result in non trivial logical
operations on the code subspace.}
 
\begin{figure}
\begin{center}
\includegraphics[scale=0.35]{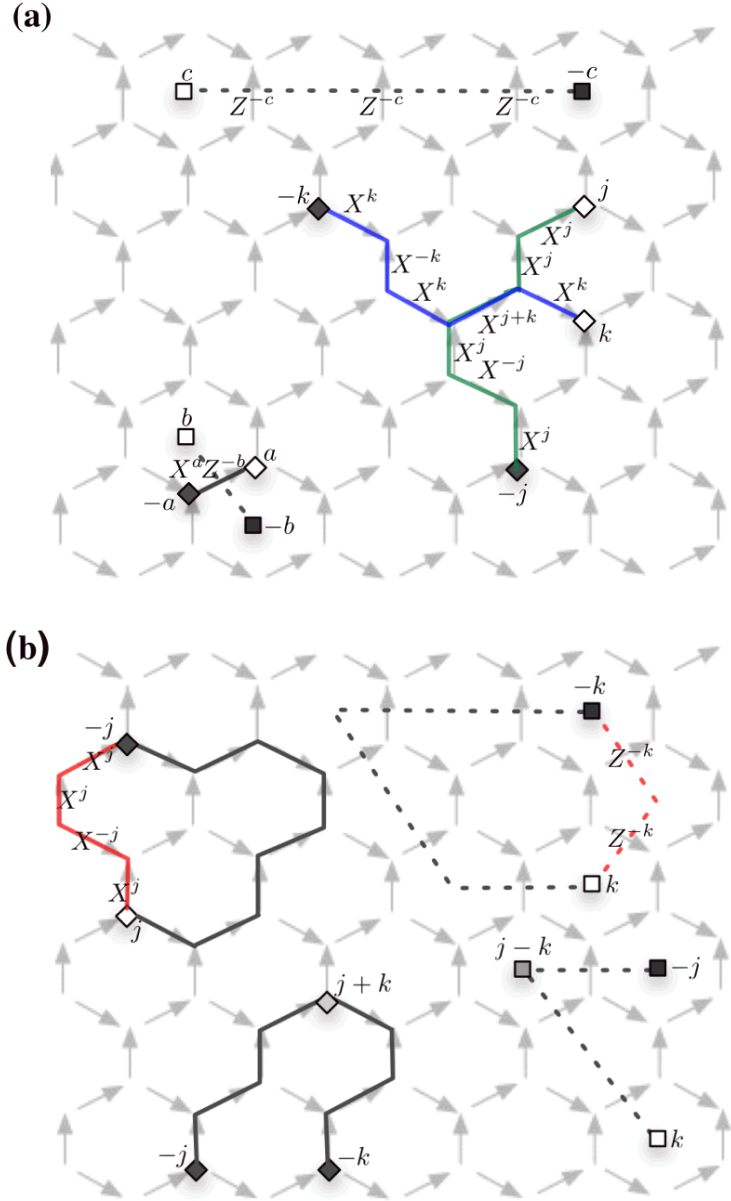}
\caption{\label{fig:2}  Quasi-particle excitations on a honeycomb 
cellulation.    (a)  Excitations appear in particle anti-particle pairs.  
Charges(anti-charges) appear as boundaries on vertices represented by 
open(filled) diamonds, and fluxes(anti-fluxes) as boundaries on the faces 
represented by open(filled) squares.  The total charge and flux of any pair 
is zero.  Shown is a flux pair $\ket{(0,c);(0,-c)}$, charge pairs 
$\ket{(j,0);(j,0)}, \ket{(k,0);(-k,0)}$ and a bound state of charge and 
flux pairs $\ket{(a,b);(-a,-b)}$.  Notice that strings of the same or 
different types are allowed to intersect.  (b)  Fusion of quasi-particles.  
The upper two diagrams illustrate corrective procedures to annihilate charge 
and flux excitations.  The lower two diagrams illustrate the fusion rules 
$\ket{(j,0)}\times\ket{(k,0)}=\ket{(j+k,0)}$ and 
$\ket{(0,j)}\times\ket{(0,-k)}=\ket{(0,j-k)}$.}
\end{center}
\end{figure}

Pauli-group elements local to a single edge of $\Gamma$
produce dyons of the topological order in particle anti-particle pairs.
To see this, note that
the operator $X^a_{e}$ acting at edge 
$e=[v_1,v_2]$ creates a pair of boundaries on the vertices, one with charge 
$a$ at $v_1$ and other 
with charge $d-a$ at $v_2$.  We name the charge $d-a$ particle an 
anti-charge to $a$.  Similarly, the operator $Z^b_e$ creates 
quasi-particles located on the two faces $f_1$ and $f_2$ that share the 
edge $e$ on their boundaries.  Let face $f_1$ be the face with opposite 
orientation to $e$.  Then the flux at $f_1$ is $b$ and the anti-flux at 
$f_2$ has the value $d-b$.  A product operator $X^aZ^{-b}$ acting on edge 
$e$ creates the dyon $(a,b)$ with charge $a$ at vertex $v_1$ and flux $b$ 
at face $f_1$ (see Fig. \ref{fig:2}a).  When it might be
clear from context,
we will drop the particle location labels $(v,f)$,
e.g. particle 
anti-particle pairs might be written as
$\ket{(a,b);(-a,-b)}$.  The mass of a dyon is given by the expectation value: 
$m_{a,b}=\bra{(a,b)}H\ket{(a,b)}-E_0=2U(1-\mbox{Re}[\xi^{a}])
+2h(1-\mbox{Re}[\xi^b])$, 
where $E_0$ is the vacuum energy.  The energy to create a particle 
antiparticle pair is twice this value.  


Prior work in continuum field theory has considered dyon excitations in which charges
and fluxes take values in $\mathbb{Z}/d\mathbb{Z}$.  The
interactions described by a $\mathbb{Z}/d\mathbb{Z}$
gauge theory are 
completely characterized by the following rules \cite{Propitius}.
\begin{equation}
\ket{(a,b;(v,f))}\times \ket{(a',b';(v,f))}=\ket{(a+a',b+b';(v,f))}
\label{fusion}
\end{equation}
\begin{equation}
\mathcal{R}\ket{(a,b;(v,f))}\ket{(a,b;(v',f'))}=
\xi^{ab}\ket{(a,b;(v,f))}\ket{(a,b;(v',f'))}
\label{braid1}
\end{equation}
\begin{equation}
\begin{array}{lll}
&&\mathcal{R}^2\ket{(a,b;(v,f))}\ket{(a',b';(v',f'))}=\\
&&\quad\quad\xi^{(a'b+b'a)}\ket{(a,b;(v,f))}\ket{(a',b';(v',f'))}
\end{array}
\label{braid2}
\end{equation}
\begin{equation}
\mathcal{C}\ket{(a,b;(v,f))}=\ket{(-a,-b;(v,f))}
\label{cong}
\end{equation}
\begin{equation}
T\ket{(a,b;(v,f))}=\xi^{ab}\ket{(a,b;(v,f))}.
\label{rot}
\end{equation}
We next review these rules and argue that the dyonic excitations
of our Hamiltonian satisfy them.

The first relation is the fusion rule for particles occupying 
the same location 
where addition is performed modulo $d$.  In the context of our model this 
rule follows from the additivity of boundaries of one chains.  Indeed, it 
is the ability to annihilate particle anti-particle pairs by choosing a 
trivial cycle on $\Gamma$ or $\tilde{\Gamma}$ that makes correction of 
local errors possible (see Fig. \ref{fig:2}b).
The next two rules describe the action of the monodromy operator 
$\mathcal{R}$ which performs a counterclockwise exchange of one particle 
with another.  The quantum state of $n$ indistinguishable particles 
residing on a surface belongs to a Hilbert space that transforms as a 
unitary representation of the braid group $B_n$.  If we order the positions 
of the particles $\{(v_j,f_j)\}_{j=1}^n$, then the $n-1$ generators of 
$B_n$ correspond to the monodromy operator $\mathcal{R}$ acting on the 
particle pairs in the locations 
$\{(v_j,f_j),(v_{j+1},f_{j+1})\}_{j=1}^{n-1}$.  For a 
$\mathbb{Z}/d\mathbb{Z}$ gauge 
theory, the irreducible unitary representation 
of $B_n$ is one dimensional, meaning the particles are abelian anyons.  
Notice that the definition of the monodromy operator involves orientation 
of the path taken during particle exchange.  For a non orientable surface, 
$\mathbb{Z}/d\mathbb{Z}$ statistics for 
$d>2$ are not allowed because the clockwise 
trajectory of particle around another is not uniquely defined whereas the 
phases $\xi,\xi^{-1}$ are distinguishable except for $d=2$. 

\begin{figure}
\begin{center}
\includegraphics[scale=0.45]{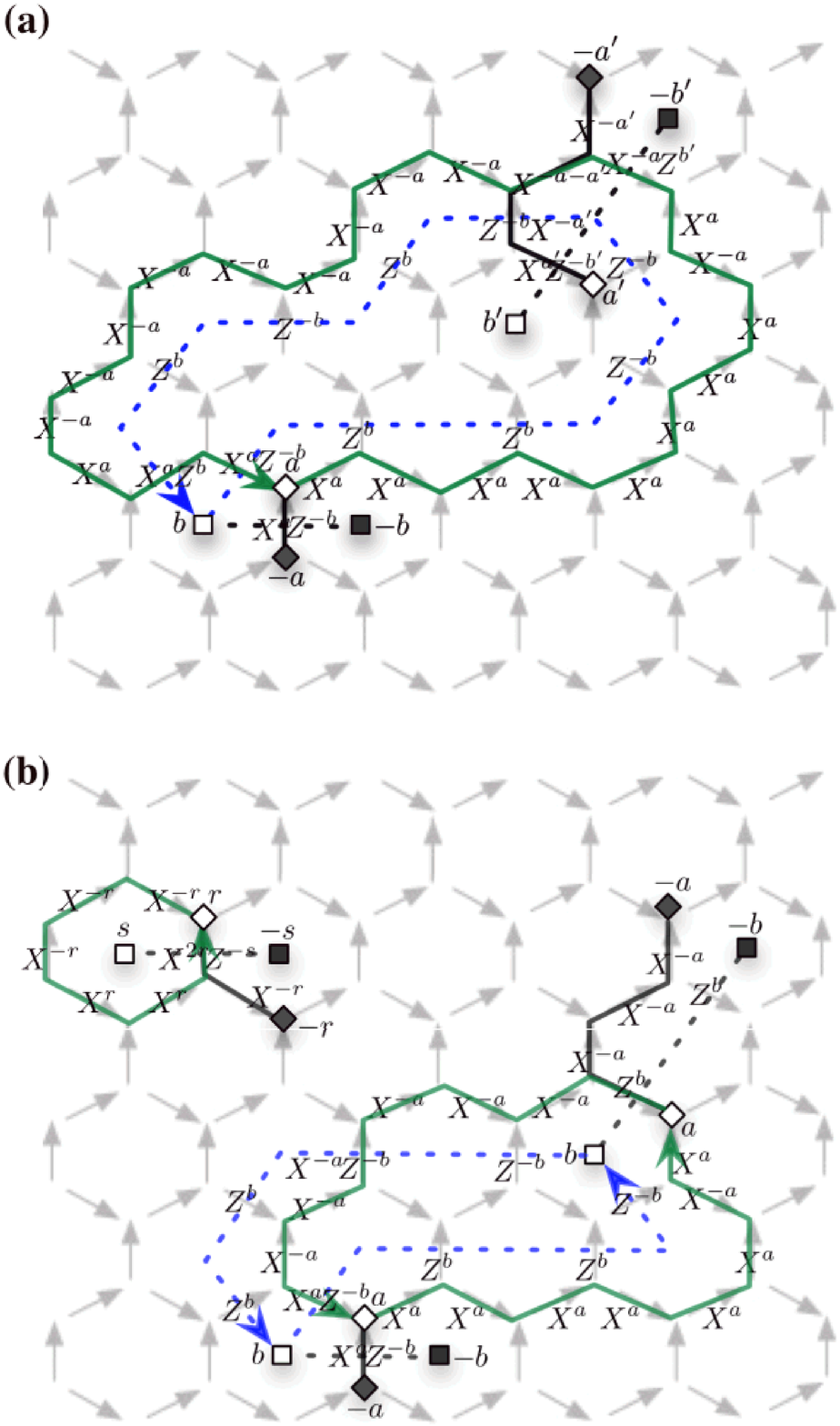}
\caption{\label{fig:3}  Braid relations.  (a)  Counterclockwise braiding 
of the dyon $(a,b)$ around the dyon $(a',b')$:  
$\mathcal{R}^2\ket{(a,b)}\ket{(a',b')}=
\xi^{(a'b+b'a)}\ket{(a,b)}\ket{(a',b')}$.  (b)  Counterclockwise exchange of 
identical dyons:  $\mathcal{R}\ket{(a,b);(a,b)}=\xi^{ab}\ket{(a,b);(a,b)}$.  
In the upper left hand side of the surface is shown the counterclockwise 
winding of the charge component of a dyon $(r,s)$ about its flux component 
generating an Aharanov-Bohm phase according to 
$T\ket{(r,s)}=\xi^{rs}\ket{(r,s)}$.}
\end{center}
\end{figure}

The braiding of one dyon around another is shown in Fig. \ref{fig:3}.  Here 
we begin with a state of two dyonic particle anti-particle pairs: 
$\ket{\Psi}=\ket{(a,b);(-a,-b)}\ket{(a',b');(-a',-b')}$ in distinct 
locations on the surface.  The mutual statistics are determined by winding 
one dyon, $(a,b)$ around the other $(a',b')$ in a counterclockwise sense.  
This action is described by the square of monodromy operator $\mathcal{R}$, 
which exchanges two particles in a counterclockwise sense. A non trivial 
phase is accumulated under the action of $\mathcal{R}^2$ because the 
closed loop string operators that wind $(a,b)$ collide with the strings 
connected the dyon $(a',b')$ with its anti-particle.  In the example 
shown in Fig. \ref{fig:3}a, the strings intersect at two locations where 
we have the operators $Z^{-b}X^{-a'}=\xi^{a'b}X^{-a'}Z^b$ and 
$X^{-a}Z^{b'}=\xi^{b'a}Z^{-b'}X^{-a}$.  Rewriting these operators with 
the action of the closed strings (Pauli operators with unprimed powers) 
first has the advantage that the closed strings act trivially 
provided that there are no other quasi-particles inside the closed loops.  
Hence we have that $\mathcal{R}^2\ket{\Psi}=\xi^{(a'b+b'a)}\ket{\Psi}$.  
The preceding example illustrated the Aharanov phase accumulated when 
winding one charge around a flux along a trajectory that was local, i.e. 
did not explore the global properties of the surface.  Were the flux 
absent, then the path would be homotopic to a point.  One can also define 
this phase for trajectories that explore the global properties of the 
surface, but can be continously deformed to a process where one anyon wraps 
around another.  On a torus, for example, the following process traces out 
non trivial cycles for the charges and fluxes.  Represent the torus as a 
square with opposite sides identified, labelling
the axes of the square 
$x_1,x_2$.  Pick a non trivial cycle along the $x_1$ direction of $\Gamma$, 
and call it $P_1$.  
Similarly, pick a non-trivial cycle along the $x_2$ 
direction of the dual $\tilde{\Gamma}$ and call it $P_2$.  To obtain the 
exchange statistics, first create the dyonic particle antiparticle pair 
$\ket{\Psi}=\ket{(a,b;(v_1,f_1));(-a,-b;(v_2,f_2))}$ out of the vacuum 
state $\ket{\Psi_g}$.  Wind the charge $a$ around $P_1$ so that it 
annihilates with its anticharge partner at site $v_2$.  Next wind the 
flux $b$ around $P_2$ so that it annihilates with its antiflux partner at 
face $f_2$.  Create another dyonic particle antiparticle pair 
$\ket{\Psi'}=\ket{(a,b;(v_2,f_2));(-a,-b;(v_1,f_1))}$ with particle 
antiparticle positions reversed relative to $\ket{\Psi}$.  Wind charge 
$a$ around $P_1$ in the opposite direction to the first winding so that 
it annihilates with the anticharge at site $v_1$ and likewise, flux 
$b$ along $P_2$ in the opposite direction so that it annihilates with the 
antiflux at face $f_2$.  These four trajectories cross at one edge $e$ and 
the action on the state (for one choice of edge orientation) is 
$\ket{\Psi_g}\rightarrow Z_e^{-b}X_e^{-a}Z_e^bX_e^a\ket{\Psi_g}=
\xi^{ab}\ket{\Psi_g}$.  If we embed the torus in $\mathbb{R}^3$, then the 
worldlines described by intersecting strings in the above process are 
equivalent under ambient isotopy to linked world lines on the plane 
which describe winding the charge $a$ around the flux $b$.   

Identical quasi-particle statistics are determined by exchanging one 
dyon $(a,b)$ counterclockwise with another.  Such a process is depicted in 
Fig. \ref{fig:3}b.  The action on the state 
$\ket{\Psi}=\ket{(a,b)}\ket{(a,b)}$ can be computed by annihilating 
particle-antiparticle pairs after exchange, creating them again, and 
comparing the resultant state with the initial state $\ket{\Psi}$.  We can 
annihilate the charges on the left side first.  Reversing the order of the 
operator that created the dyon there, we have 
$X^aZ^{-b}=\xi^{ab}Z^bX^a$ and the charges are annihilated by applying 
$Z^{b}$.  Similarly, the charges on the right side are annihilated by 
applying a string of $Z^{-b}$ operators.  Finally, the fluxes are annihilated 
by applying $X^a$ or $X^{-a}$ along the remaining two connected 
strings.  The action on the wavefunction is then 
$\mathcal{R}\ket{(a,b)}\ket{(a,b)}=\xi^{ab}\ket{(a,b)}\ket{(a,b)}$.

The particle conjugation operator $\mathcal{C}$ in Eq. \ref{cong} reverses the 
sign of all the particles.  This is realized in our microscropic spin model 
by reversing the orientation of the all the edges on the cellulation.   
Finally, the operation $T$ in rule \ref{rot} rotates the charge component of a 
dyon around its own flux, generating an Aharanov-Bohm phase in the process.  
This is illustrated in Fig. \ref{fig:3}b.  Here the charge component of the 
dyon $(r,s)$ is wrapped around its flux component in a counterclockwise sense.
During this operation, there is a collision at the edge where the dyon was 
created.  Rewriting the operation on the edge as 
$X^{2r}Z^{-s}=\xi^{rs}X^rZ^{-s}X^r$ so that loop operation about boundary 
of the face $f$ acts trivially first, we have that 
$T\ket{r,s(v,f)}=\xi^{rs}\ket{r,s(v,f)}$.  

\subsection{Measuring statistical phases}

\begin{figure}
\begin{center}
\includegraphics[scale=0.22]{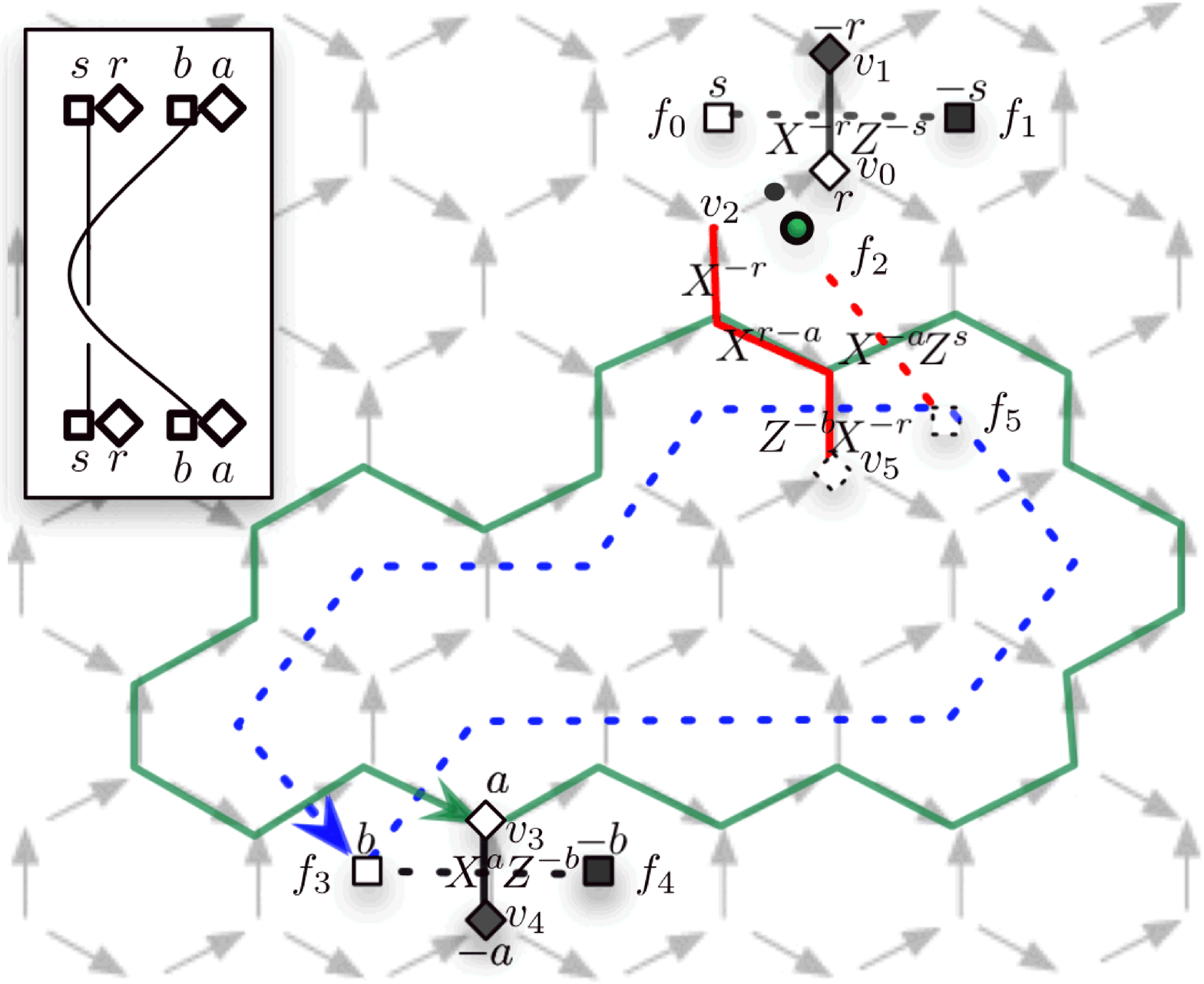}
\caption{\label{fig:4}  Protocol for measuring quasi-particle statistics.  
The green circle represents an ancillary particle which performs conditional 
gate operations on the qudit residing on edge $e=[v_2,v_0]$.  The red lines 
indicate operations which are done adiabatically with respect to the energy 
gap $\Delta E$.  The inset is a simplified space-time diagram of the braid.}
\end{center}
\end{figure}

In any physical construction of a Hamiltonian that admits topologically 
ordered states it will be important to verify the predicted properties.  
One, albeit crude, observable is to measure the energy gap
from a ground state to a first excited state.  
This could be done by probing linear response of the ground states to a 
perturbing field that generates local unitary operation at a frequency 
$\omega_F$.  For a system with the internal Hamiltonian Eq. \ref{totHam}, 
the expected resonant absorption occurs at frequencies 
$\omega_F=2m_{a,b}/\hbar$.
However, as a witness to topological order, this measure is 
not sufficient because there could be another spin Hamiltonian 
with equal gap that does not  possess topologically invariant correlations 
functions.  Another more convincing probe would be to directly compute the 
statistical phases in Eq. \ref{braid2}.  Operationally, this should be done 
by measuring both the phase $\phi_{\tau}$ accumulated when one particle 
$(a,b)$ wraps around another $(r,s)$ and the phase $\phi_{\bf 1}$ when the 
particle $(a,b)$ traces out the same path in configuration space but does 
not enclose the particle $(r,s)$.  The phase difference 
$\phi_{\tau}-\phi_{\bf 1}=\phi_{\rm top}$ subtracts out dynamical phases 
and Berry's phases, leaving only topological information.  We sketch an 
algorithm for computing this phase using operations in accordance with the 
two complex illustrated in Fig. \ref{fig:4}.  Adaptation to other 
cellulations is straightforward.

\begin{enumerate}
\item\label{1} Beginning from a ground state $\ket{\Psi (0)}$, prepare a state 
with two particle anti-particle pairs in disjoint regions of the surface:  
\[
\begin{array}{lll}
\ket{\Psi(1)}&=&\ket{(a,b;(v_3,f_3));(-a,-b;(v_4,f_4))}\\
& &\ket{(r,s;(v_0,f_0));(-r,-s;(v_1,f_1))}.
\end{array}
\]

\item\label{2} Prepare an ancillary qubit $a$ in the state 
$\ket{+_x}_a=\frac{1}{\sqrt{2}}(\ket{0}_a+\ket{1}_a)$ and use this qubit to 
perform the controlled unitary operation 
$\wedge_1(X_e^{-r}Z_e^{s})=\ket{0}_a\bra{0}\otimes {\bf 1}_d+
\ket{1}_a\bra{1}\otimes X_e^{-r}Z_e^{s}$ (with $(r,s)\neq (0,0)$) on the 
qudit residing on the edge $e=[v_2,v_0]$.  Measure the ancilla in the 
$\hat{x}$ basis and record the result $m=\pm 1$.  The resultant state is 
$\ket{\Psi(2)}=\frac{1}{\sqrt{2}}(\ket{\Psi(1)}+
(-1)^{m} X_e^{-r}Z_e^{s}\ket{\Psi (1)})$, where
\[ 
\begin{array}{lll}
X_e^{-r}Z_e^{s}\ket{\Psi(1)}&=&\ket{(a,b;(v_3,f_3));(-a,-b;(v_4,f_4))}\\
& &\ket{(r,s;(v_2,f_2));(-r,-s;(v_1,f_1))}
\end{array}
\]
is orthogonal to $\ket{\Psi(1)}$.

\item\label{3}  Use a sequence of local spin operations to drag the dyon at 
location $(v_2,f_2)$ to the location  
$(v_5,f_5)$.  These operations should be done adiabatically, i.e. they should 
be done using localized control fields with frequency components much smaller 
than the minimum gap energy $\Delta E$.  In this way no new particles will be 
created, only the component of the wavefunction with the dyon located at 
$(v_2,f_2)$ will be changed.  Instead of using control 
fields to perform local spin operations, another possibility is to slowly decrease 
the values of $U$ and $h$ on the vertices and faces in the path from $(v_2,f_2)$ to 
$(v_5,f_5)$ so that it is energetically favorable for the dyon to follow 
this path.   The resultant state is:
$\ket{\Psi(3)}=\frac{1}{\sqrt{2}}(\ket{\Psi(1)}+(-1)^{m}\ket{\Psi '})$, where
\[ 
\begin{array}{lll}
\ket{\Psi '}&=&\ket{(a,b;(v_3,f_3));(-a,-b;(v_4,f_4))}\\
& &\ket{(r,s;(v_5,f_5));(-r,-s;(v_1,f_1))}.
\end{array}
\]

\item
\label{4} Braid the dyon $(a,b;(v_3,f_3))$  in a counterclockwise sense around 
the location $(v_5,f_5)$ such that it returns to location $(v_3,f_3)$.  The 
state is now:  
$\ket{\Psi(4)}=\frac{1}{\sqrt{2}}(\ket{\Psi(1)}+
(-1)^{m} \xi^{(sa+rb)}\ket{\Psi '})$. 

\item\label{5} Perform the inverse of the operations in step \ref{3}, again 
insuring that no new quasi-particles are created during the process.  The 
resulting state is:  
$\ket{\Psi(5)}=\frac{1}{\sqrt{2}}(\ket{\Psi(1)}+
(-1)^{m} \xi^{(sa+rb)}e^{i\chi}X_e^{-r}Z_e^{s}\ket{\Psi(1)})$, where we have 
included $\chi$, the sum of dynamical and Berry's phases that may have 
accumulated during steps \ref{2}-\ref{4}.

\item
\label{6}  Reprepare the ancilla in the state $\ket{+_x}_a$ and perform the 
controlled unitary operation $\wedge_1((-1)^{m}Z_e^{-s}X_e^r)$.  Measure the 
qubit in the $\hat{x}$ basis.  The expectation value is:
\[
\begin{array}{lll}
\langle \sigma^x_a\rangle_{\tau}&=&\frac{1}{2}\Big(\cos(\chi+\phi_{\rm top})+\\
& &\delta_{2r,0}\delta_{2s,0}\cos(\chi+\phi_{\rm top}-2\pi rs/d)\Big),
\end{array}
\]
where $\phi_{\rm top}=2\pi (sa+rb)/d$.  
 
\item\label{7} Repeat steps \ref{1}-\ref{6} but measure the ancilla in the 
$\hat{y}$ basis.  The expectation value is:
\[
\begin{array}{lll}
\langle \sigma^y_a\rangle_{\tau}&=&\frac{1}{2}\Big(\sin(\chi+\phi_{\rm top})-\\
& &\delta_{2r,0}\delta_{2s,0}\sin(\chi+\phi_{\rm top}-2\pi rs/d)\Big),
\end{array}
\]

\item Perform a similar experiment but this time using a trivial braiding 
operation, i.e. perform the steps
in the order (\ref{1},\ref{2},\ref{4},\ref{3},\ref{5},\ref{6},\ref{7}) so 
that the braid is contractible.  Then the expectation values are
\[
\langle \sigma^x_a\rangle_{\bf 1}=\frac{1}{2}\Big(\cos\chi+\delta_{2r,0}
\delta_{2s,0}\cos(\chi-2\pi rs/d)\Big),
\]
\[
\langle \sigma^y_a\rangle_{\bf 1}=
\frac{1}{2}\Big(\sin\chi-
\delta_{2r,0}\delta_{2s,0}\sin(\chi-2\pi rs/d)\Big).
\]

\item Compute the topological phase $\phi_{\rm top}$ from an ensemble average 
obtained by repeated measurements on identically prepared systems.

\end{enumerate}  

As a simple example, consider the computation of the mutual statistics of 
charge and a flux for $d=2$.  Setting $(r,s)=(0,1)$ and $(a,b)=(1,0)$, the 
expected measurement results are 
$\langle \sigma^x_a\rangle_{\tau}=\frac{1}{2}\cos(\chi+\phi_{\rm top})$,  
$\langle \sigma^y_a\rangle_{\tau}=0$, 
$\langle \sigma^x_a\rangle_{\bf 1}=\frac{1}{2}\cos\chi$,  
$\langle \sigma^y_a\rangle_{\bf 1}=0$.  If desired, the 
phase $\chi$ could be engineered to vary in a controlled manner over 
different trials in order to improve the visibility of the phase shift 
$\phi_{\rm top}$.  For $d>2$ it is always possible to choose the probe 
dyon such that $\delta_{2r,0}\delta_{2s,0}=0$.  In this case, 
$\phi_{\rm top}$ is estimated by finding the closest solution to 
$e^{i\phi_{\rm top}}=(\langle \sigma^x_a\rangle_{\tau}+i
\langle \sigma^y_a\rangle_{\tau})/
( \langle \sigma^x_a\rangle_{\bf 1}+i\langle \sigma^y_a\rangle_{\bf 1})$.

\section{Conclusions}
We have proven the existence of a microscopic spin model that provides for topologically 
protected qudit encodings.  This model describes
a $\mathbb{Z}/d\mathbb{Z}$ gauge theory with abelian 
charge/flux dyons as excitations.
The construction is quite general, allowing for arbitrary 
cellulations of an orientable surface
and encoding qudits with any finite number of levels.  Suggested 
adaptations to the standard 
spin models using ancilla for in place stabilizer checks could prove 
advantageous in any physical 
implementation of such codes.  
Moreover, with some limited degree 
of local control it is possible to measure the anyonic statistical phases.  
Given that these properties 
are difficult to measure in quantum Hall systems, this could provide a novel 
probe of topological order.

\subsection{Acknowledgments}
GKB appreciates helpful conversations with Xiao-Gang Wen, Bei-Lok Hu, 
and John Preskill.  Part of this work was completed at the Kavli 
Institute for Theoretical Physics 2006 Workshop on Topological Phases 
and Quantum Computation.  This research was supported in part by the 
Austrian Science Foundation and the
National Science Foundation under Grant No. PHY99-07949.


\end{document}